\documentclass[11pt,onecolumn]{IEEEtran}
\usepackage{amsthm}
\usepackage{times,amssymb,amsmath,amsfonts,float,nicefrac,color,bbm,mathrsfs,caption,float}
\usepackage{algorithm,enumerate,multirow,caption,tikz,graphicx}
\usepackage[mathscr]{eucal}
\usepackage{sidecap}
\usepackage{algpseudocode}
\usepackage{verbatim}
\usepackage{epstopdf}
\usepackage{textcomp}
\usetikzlibrary{shapes,arrows}
\usepackage{stmaryrd}
\usepackage{mathabx}
\usepackage[noadjust]{cite}
\usepackage{booktabs}
\usepackage[normalem]{ulem}
\usepackage[top=1in,bottom=1in,left=1in,right=1in]{geometry}
\allowdisplaybreaks

\newcommand{\RN}[1]{%
  \textup{\expandafter{\romannumeral#1}}%
}

\newcommand{\Mod}[1]{\ (\textup{mod}\ #1)}
\newcommand\remove[1]{}

\allowdisplaybreaks
\usepackage{enumitem}
\setlist[enumerate]{leftmargin=*}

\newtheorem{theorem}{Theorem}
\newtheorem{definition}{Definition}

\newtheorem{lemma}{Lemma}

\newtheorem{remark}{Remark}

\newtheorem{cnstr}{Construction}

\newcommand{\cC}{\mathcal{C}}

\newcommand{\cF}{\mathcal{F}}

\newcommand{\cR}{\mathcal{R}}
\newcommand{\cS}{\mathcal{S}}

\DeclareMathOperator{\ce}{ce}
\DeclareMathOperator{\co}{co}

\usepackage{tikz}
\usetikzlibrary{calc}

\begin{document}
\title{Cooperative repair: Constructions of optimal MDS codes for all admissible parameters}

\author{\IEEEauthorblockN{Min Ye} \hspace*{1in}
\and \IEEEauthorblockN{Alexander Barg}}

\maketitle
{\renewcommand{\thefootnote}{}\footnotetext{

\vspace{-.2in}
 
\noindent\rule{1.5in}{.4pt}

M. Ye is with Department of Electrical Engineering, Princeton University, Princeton, NJ, email: yeemmi@gmail.com. A. Barg  is with Department of Electrical and Computer Engineering and Institute for Systems Research, University of Maryland, College Park, MD 20742 and IITP RAS, Moscow, Russia, email: abarg@umd.edu. Research partially supported by NSF grants CCF1422955 and CCF1618603.
}
\renewcommand{\thefootnote}{\arabic{footnote}}
\setcounter{footnote}{0}

\begin{abstract}
Two widely studied models of multiple-node repair  in distributed storage systems are {\em centralized repair} and {\em cooperative  repair}. The centralized model assumes that all the failed nodes are recreated in one location, while the cooperative one stipulates that the failed nodes may communicate but are distinct, and the amount of data exchanged between them is included in the repair bandwidth.

As our first result, we prove a lower bound on the minimum bandwidth of cooperative repair. We also show that the cooperative model is stronger than the centralized one, in the sense that any MDS code with optimal repair bandwidth under the former model also has optimal bandwidth under the latter one. These results were previously known under the additional ``uniform download'' assumption, which is removed in our proofs.

As our main result, we give explicit constructions of MDS codes with optimal cooperative repair for all possible parameters. More precisely, given any $n,k,h,d$ such that $2\le h \le n-d\le n-k$ we construct $(n,k)$ MDS codes over the field $F$ of size $|F|\ge (d+1-k)n$ that can optimally repair any $h$ erasures from any $d$ helper nodes. The repair scheme of our codes involves two rounds of communication. In the first round, each failed node downloads information from the helper nodes, and in the second one, each failed node downloads additional information from the other failed nodes. This implies that our codes achieve the optimal repair bandwidth using the smallest possible number of rounds.
\end{abstract}

\section{Introduction}\label{section:intro}
\subsection{Centralized and cooperative repair models} The problem considered in this paper is motivated by the distributed nature of the system wherein the coded data is distributed across a large number of physical storage nodes. When some storage nodes fail, the repair task performed by the system relies on communication between individual nodes, which introduces new challenges in the code design. Coding schemes that address these challenges are known under the name of {\em regenerating codes}, a concept that was isolated and studied in the work of Dimakis et. al. \cite{Dimakis10}. In paper \cite{Dimakis10} the authors suggested a new metric that has a bearing on the overall efficiency
of the system, namely, the {\em repair bandwidth}, i.e., the amount of data communicated between the nodes in the process of repairing failed nodes.  Most works on this class of codes assume that the information is protected with 
Maximum Distance Separable (MDS) codes which provide the optimal tradeoff between failure tolerance and storage overhead. Paper \cite{Dimakis10} also gave a lower bound on the minimum repair bandwidth of MDS codes, 
known as the {\em cut-set bound}. Code families that achieve this bound with equality are said to have the {\em optimal repair property}. 
Constructions of optimal-repair MDS codes (also known as {\em minimum storage regenerating}, or MSR codes) were proposed in 
\cite{Rashmi11,Tamo13,Ye16,Sasid16,Ye16a,Tamo17RS}.

To encode information with an MDS code, the original file is divided into $k$ information blocks viewed as vectors over a finite field $F$. The encoding procedure then finds $r=n-k$ parity blocks, also viewed as vectors over $F$, which together with the information blocks form a codeword of a code of length $n$. The $n$ blocks of the codeword are stored on $n$ different storage nodes. Motivated by this model, we also refer to the coordinates of the codeword as nodes. The task of node repair therefore
amounts to erasure correction with the chosen code, and the special feature of the erasure correction problem arising from
the distributed data placement is the constraint on the repair bandwidth involved in the repair procedure.

Most studies of MDS codes with optimal repair bandwidth in the literature are concerned with a particular subclass of codes known as MDS {\em array codes} \cite{Blaum98}. An $(n,k,l)$ MDS array code over a finite field $F$ is formed of $k$ information nodes and $r=n-k$ parity nodes with the property that the contents of any $k$ out of $n$ nodes suffices to recover the codeword.   Every
node is a column vector in $F^l,$ reflecting the fact that the system
views a large data block stored in one node as one coordinate of the codeword.
The parameter $l$ that determines the dimension
of each node is called {\em sub-packetization}.

While originally the repair problem was confined to a single node failure, studies into regenerating codes
have expanded into the task of repairing multiple erasures. The problem of repairing multiple erasures comes in
two variations.
One of them is the {\em centralized model}, where a single data center is responsible for the repair of all the failed nodes \cite{Cadambe13,Ye16,Rawat16a,Wang17,Zorgui17,Ye17,Zorgui18}, and the other is the {\em cooperative model}, where the failed nodes may communicate but are distinct, and the amount of data exchanged between them is included in the repair bandwidth \cite{Kermarrec11,Shum13,Li14,Shum16}.
The cut-set bounds on the repair bandwidth for multiple erasures under these two models were derived in \cite{Cadambe13} and \cite{Shum13} respectively.

Let $\cF\subset [n], |\cF|=h$ and $\cR\subseteq[n]\backslash \cF, |\cR|=d$ be the sets of indices of the failed nodes and
the helper nodes, respectively, where we use the notation $[n]:=\{1,2,\dots,n\}.$
Informally speaking, under the {centralized model}, repair proceeds by downloading $\beta_j,j\in\cR$ symbols of $F$ from {each of} the helper nodes {$C_j,j\in\cR$}, and computing the values of the failed nodes. It is assumed that the repair is performed
by a data center having access to all the downloaded information, and so the {\em repair bandwidth} equals $\beta_\cF(\cR)=\sum_{j\in\cR}\beta_j$. The variation introduced by the {cooperative model} does not include the data center, and so the repair bandwidth includes not only the information downloaded from the helper nodes but also the information exchanged between the 
failed nodes in the repair process. In other words, under the centralized model, each failed node has access to all the data downloaded from the helper nodes, while under the cooperative model, each failed node only has access to its own downloaded data.

\subsection{Formal statement of the problems}
Consider an $(n,k,l)$ MDS array code $\cC$ over a finite field $F$ and let $C\in \cC$ be a codeword. 
We write $C$ as $(C_1,C_2,\dots,C_n)$,
where $C_i=(c_{i,0},c_{i,1},\dots,c_{i,l-1})^T\in F^l, i=1,\dots, n$ is the $i$th coordinate of $C$. 
The node repair models can be formalized as follows.
\begin{definition}[Centralized model]
Let $\cF$ and $\cR$ be the sets of failed and helper nodes, and suppose that $|\cF|=h\le r$ and $|\cR|=d\ge k.$
We say that the failed nodes $\{C_i,i\in\cF\}$ can be repaired from the helper nodes $\{C_j,j\in\cR\}$
by downloading\footnote{We note the use of the application-inspired term ``download'' for evaluating the functions $f_j$ and making their values
available to the failed nodes. This term is used extensively throughout the paper.} $\beta_{\cF}(\cR)$ symbols of $F$ if there are
$d$ numbers $\beta_j, j\in\cR$, $d$ functions $f_j: F^l\to F^{\beta_{j}}, j\in\cR,$
and $h$ functions $g_i: F^{\sum_{j\in\cR}\beta_{j}}\to F^l, i\in\cF$
such that
\begin{enumerate}
\item for every $i\in \cF$ and every $C\in\cC$
$$
C_i=g_i(\{f_j(C_j),j\in\cR \}),
$$
\item
$$
\sum_{j\in\cR}\beta_j=\beta_{\cF}(\cR).
 $$
\end{enumerate}
\end{definition}

Under the cooperative model, the repair process is divided into two rounds. In the first round, each failed node downloads data from the helper nodes, and in the second round, the failed nodes exchange data among themselves (namely, each failed node downloads data from the other failed nodes).

\begin{definition}[Cooperative model] \label{def:op} In the notation of the previous definition, we assume two rounds
of communication between the nodes. In the first round, each failed node $C_i,i\in\cF$ downloads a vector $f_{ij}(C_j)$ from each helper node $C_j,j\in\cR,$ and in the second round, each failed node $C_i,i\in\cF$ downloads a vector 
$f_{ii'}(\{f_{i'j}(C_j),j\in\cR \})$ from each of the other failed nodes $C_{i'},i'\in\cF\setminus\{i\}$.
We require that each failed node $C_i,i\in\cF$ can be recovered from its own downloaded data
$f_{ij}(C_j),j\in\cR$ and $f_{ii'}(\{f_{i'j}(C_j),j\in\cR \}),i'\in\cF\setminus\{i\}$.
The amount of downloaded data in this two-round repair process is
$$
\sum_{i\in\cF}\Big(\sum_{j\in\cR} \dim_F \big( f_{ij}(C_j) \big) 
+ \sum_{i'\in\cF\setminus\{i\}} \dim_F \big( f_{ii'}(\{f_{i'j}(C_j),j\in\cR \}) \big) \Big),
$$
where $\dim_F(\cdot)$ is the dimension of the argument expressed as a vector over $F.$
\end{definition}

This definition may look somewhat restrictive in the part where the communication is constrained to only two rounds. 
Indeed, in the definition proposed in \cite{Shum13}, the repair process may include an arbitrary number $T$ of communication rounds. 
However, in this paper we show that it suffices to consider $T=2$ to construct codes with optimal repair bandwidth for all
possible parameters, and therefore we rely on the above definition, which also leads to simplified notation. At the same time, it may be that for other problems of cooperative repair, such 
as optimal-access repair or others, more than two rounds are in fact necessary.

Given a code $\cC$, define $N_{\ce}(\cC,{\cF},{\cR})$ and $N_{\co}(\cC,{\cF},{\cR})$ as the smallest number of symbols of $F$ one needs to download  in order to recover the failed  
nodes $\{C_i,i\in{\cF}\}$ from the helper nodes $\{C_j,j\in{\cR}\}$ under the centralized model and the cooperative model, respectively.
The repair bandwidth of the code is defined as follows.
\begin{definition}[Repair bandwidth]
Let $\cC$ be an $(n,k,l)$ {MDS} array code over a finite field $F$.
  The \emph{$(h,d)$-repair bandwidth} of the code $\cC$ under centralized/cooperative repair model is given by 
  \begin{equation}\label{eq:beta}
\begin{aligned}
\beta_{\ce}(h,d):=\max_{|{\cF}|=h,|{\cR}|=d, {\cF}\bigcap{\cR}=\emptyset} N_{\ce}(\cC,{\cF},{\cR}),\\
\beta_{\co}(h,d):=\max_{|{\cF}|=h,|{\cR}|=d, {\cF}\bigcap{\cR}=\emptyset} N_{\co}(\cC,{\cF},{\cR}).
\end{aligned}
  \end{equation}
\end{definition}

As already mentioned, the quantity $\beta(h,d)$ satisfies a general lower bound. In the next theorem we collect
results from several papers that establish different versions of this result.
\begin{theorem}[Cut-set bound \cite{Dimakis10,Cadambe13,Shum13}, this paper] \label{def:csb}
Let $\cC$ be an $(n,k,l)$ MDS array code. For any two disjoint subsets ${\cF},{\cR}\subseteq[n]$ such that $|{\cF}|\le r$ and $|{\cR}|\ge k,$ we have the following inequalities:
\begin{align}
N_{\ce}(\cC,{\cF},{\cR}) & \ge \frac{|{\cF}||{\cR}|l}{|{\cF}|+|{\cR}|-k}, \label{eq:csce}\\
N_{\co}(\cC,{\cF},{\cR}) & \ge \frac{|{\cF}|(|{\cR}|+|{\cF}|-1)l}{|{\cF}|+|{\cR}|-k}. \label{eq:cutset}
\end{align}
\end{theorem}

We note that in \cite{Shum13}, the bound \eqref{eq:cutset} was proved under the additional assumption that each failed node downloads the same amount of data from each helper node, and each failed node also downloads the same amount of data from each of the other failed nodes (the {\em uniform download assumption}), while our proof of \eqref{eq:cutset} in this paper does not require any additional assumptions.
A self-contained rigorous proof of \eqref{eq:cutset} is given in Section~\ref{ap:st} as a part of the proof of Theorem \ref{thm:st} below.

Inequality \eqref{eq:csce} gives the cut-set bound for the centralized model, and \eqref{eq:cutset} gives the cut-set bound under the cooperative one. For the case of a single failed node, there is no difference between the two repair models, and
these bounds coincide.

Note that although in this paper we consider only two-round cooperative repair schemes, bound \eqref{eq:cutset} holds for cooperative repair with any number of communication rounds.
 If $\beta_{\ce}(h,d)$ (resp., $\beta_{\co}(h,d)$) meets the bound \eqref{eq:csce} (resp., \eqref{eq:cutset}) with equality, i.e.,
$$
\beta_{\ce}(h,d)=\frac{hdl}{h+d-k} \quad 
\Big( \text{resp.,~~} \beta_{\co}(h,d)=\frac{h(h+d-1)l}{h+d-k} \Big),
$$
 we say that the code $\cC$ has the {\em $(h,d)$-optimal repair property} under the centralized (resp., cooperative) model.

Let us give a heuristic argument in favor of \eqref{eq:cutset} based on the cut-set bound for repairing single erasure. Let $i$ be one of the indices of the failed nodes. Suppose that all the other failed nodes $C_j,j\in\cF\setminus\{i\}$ are functional, and we need to repair $C_i$. Using either \eqref{eq:csce} or \eqref{eq:cutset} with $|\cF|=1,$ we see that $C_i$ needs 
to download at least $l/(|\cF|+|\cR|-k)$ field symbols from each of the nodes $C_j,j\in\cR\cup\cF\setminus\{i\}.$ 
Therefore each failed node $C_i,i\in\cF$ needs to download at least $(|\cF|+|\cR|-1)l/(|\cF|+|\cR|-k)$ symbols of $F$ in total. 
Thus, if \eqref{eq:cutset} is achievable with equality, then each failed node can be repaired as though all the other failed nodes were functional and available.
We note that this argument is not rigorous because the single-erasure cut-set bound is derived under a one-round repair process while the repair process under the cooperative model is divided into two rounds.

The argument in the previous paragraph also suggests that optimality of a code under cooperative repair implies 
its optimality under centralized repair.
We formalize this idea in the next theorem.
\begin{theorem}[{Cooperative model is stronger than centralized model}]\label{thm:st}
Let $\cC$ be an $(n,k,l)$ MDS array code and let ${\cF},{\cR}\subseteq[n]$ be two disjoint subsets such that $|{\cF}|\le r$ and $|{\cR}|\ge k.$ If
\begin{equation}\label{eq:of}
N_{\co}(\cC,{\cF},{\cR}) = \frac{|{\cF}|(|{\cR}|+|{\cF}|-1)l}{|{\cF}|+|{\cR}|-k},
\end{equation}
then
\begin{equation}\label{eq:xv}
N_{\ce}(\cC,{\cF},{\cR}) = \frac{|{\cF}||{\cR}|l}{|{\cF}|+|{\cR}|-k}.
\end{equation}
The statement of the theorem holds for cooperative repair schemes with any number $T\ge 2$ of communication rounds.
\end{theorem}

The statement in Theorem~\ref{thm:st} is trivially true under the uniform download assumption and in this form it was stated in \cite{Rawat16a}. In this paper we prove the theorem in Section~\ref{ap:st} under no additional assumptions.
The following arguments provide an intuitive explanation of its claim in the case of $T=2$, and they can be easily extended to any $T$.
As mentioned above, for \eqref{eq:of} to hold with equality, each failed node $C_i,i\in\cF$ should
download $l/(|\cF|+|\cR|-k)$ symbols of $F$ from each of the nodes $C_j,j\in\cR\cup(\cF\setminus\{i\})$ in the course
of the two-round repair process. Therefore, each failed node $C_i,i\in\cF$
downloads only $|\cR|l/(|\cF|+|\cR|-k)$ symbols of $F$ in total from all the helper nodes $\{C_j,j\in\cR\}.$ 
Switching to the centralized model, we observe that once these symbols are made available to one failed node, 
they are automatically available to all the other failed nodes at no cost to the bandwidth, and so
 \eqref{eq:xv} follows immediately.

\remove{
\begin{definition}[Optimal repair property]\label{def:orp}
We say that an $(n,k,l)$ MDS code $\cC$ has the \emph{$(h,d)$-optimal repair} property under the centralized model if the $(h,d)$-repair bandwidth of $\cC$ (see \eqref{eq:beta})
equals 
$$
   \beta_{\ce}(h,d)=\frac{hdl}{h+d-k},
$$
meeting the lower bound in \eqref{eq:csce} with equality.
Similarly, we say that an $(n,k,l)$ MDS code $\cC$ has the \emph{$(h,d)$-optimal repair} property under the cooperative model if the $(h,d)$-repair bandwidth of $\cC$ (see \eqref{eq:beta})
equals 
$$
   \beta_{\co}(h,d)=\frac{h(h+d-1)l}{h+d-k},
$$
meeting the lower bound in \eqref{eq:cutset} with equality.
 \end{definition}}
According to Theorem~\ref{thm:st}, MDS codes with $(h,d)$-optimal repair property under the cooperative model also have the same property under the centralized model. At the same time, it is not known how to transform optimal centralized-repair codes
into cooperative-repair codes. This might be the reason why the latter are more difficult to construct.
Indeed, while general $(h,d)$-optimal repair MDS codes for the centralized model are available in several variations \cite{Ye16,Goparaju17,Ye17}, MDS codes with the same property under the cooperative model are known only for some special values of $h$ and $d$. Specifically, the following results appeared in the literature. Paper \cite{Shum13} constructed optimal MDS codes for cooperative repair for the (trivial) case $d=k$, and \cite{Li14} presented a family of optimal MDS codes for the repair of two erasures in the regime of low rate $k/n \le 1/2$ (more precisely, \cite{Li14} constructed $(n,k)$ MDS codes with the $(2,d)$-optimal repair property for any $n,k,d$ such that $2k-3\le d\le n-2$).

Thus, prior to our work, even the existence problem of cooperative MDS codes with the $(h,d)$-optimal repair property for general values of $h$ and $d$ (apart from the two special cases mentioned above) was an open question\footnote{In \cite{Shum13}, the authors showed that the cut-set bound \eqref{eq:cutset} is achievable under the weaker ``functional repair" requirement, which does not assume that the repair scheme recovers the exact content of the failed nodes, as opposed to the more prevalent exact repair requirement considered in this paper.}.

In the rest of the paper we focus on the cooperative model, and, unless stated otherwise, all the concepts
and objects mentioned below such as the repair bandwidth, the cut-set bound, etc., implicitly assume this
model.

Our results in this work are as follows:
\begin{enumerate}
\item We give a complete solution of repairing multiple erasures for all possible parameters. More precisely, given any $n,k,h,d$ such that $2\le h \le n-d\le n-k-1$, we present an explicit $(n,k)$ MDS code with the 
$(h,d)$-optimal repair property.
We limit ourselves to the case of $d\ge k+1$ because constructions for $d=k$ were already given in \cite{Shum13}.

The size of the underlying finite field is $sn$ for all constructions, where $s:=d+1-k.$ At the same time, the sub-packetization $l$ is rather large: for $h=2$ we need to take approximately $l=s^{n(n-1)}$, while for general $d$ and $h$ it is approximately $l=s^{h\binom nh}.$ We do not know
whether this is necessary or is merely an artifact of our construction.

\item 
We prove the cut-set bound \eqref{eq:cutset} for the most general case without the uniform download assumption, and
we also show that the any MDS code that affords cooperative optimal repair is also optimally repairable under the centralized model
(see Theorem~\ref{thm:st}).
\end{enumerate}

\subsection{Organization of this paper}  
In Section~\ref{ap:st}, we prove the general versions of the cut-set bound \eqref{eq:cutset} and Theorem~\ref{thm:st} without the uniform download assumption.

In Section~\ref{sect:sl} we prove a technical lemma which forms the core of the proposed repair schemes. Various versions of this lemma will be used throughout the paper. 
Moving to the code constructions, we start with the special case of $h=2$ and $d=k+1$ to illustrate the new ideas behind the proposed code families. These results are presented in Section~\ref{sect:bdblock}. Namely, in 
Section~\ref{sect:first} we construct MDS codes $\cC_{2,k+1}^{(0)}$ that can optimally repair the first two nodes 
(or any {\em given} pair of nodes) from any $d=k+1$ helper nodes. In Section~\ref{sect:warmup}, we use this code as a building block
 to construct $(n,k)$ MDS codes $\cC_{2,k+1}$ with the $(2,d=k+1)$-optimal repair property. 
 
 In Section~\ref{sect:gend}, we deal with general values of $d, k+1\le d\le n-2$. Similarly to the above, in Section~\ref{sect:fd} we construct a code $\cC_{2,d}^{(0)}$ that supports
optimal repair of the first two nodes, and in Section~\ref{sect:rb} we use it 
 as a building block to construct MDS codes $\cC_{2,d}$ with the $(2,d)$-optimal repair property for general values of $d, k+1\le d\le n-2$.
 
In Section~\ref{sect:h} we construct $(n,k)$ MDS codes with $(h,d=k+1)$-optimal repair property for general values of $h, 2\le h\le r-1$. Following the route chosen above, in Section~\ref{sect:gh} we handle the case of repairing the first $h$ nodes while in Section~\ref{sect:lo} we extend the construction to repair any subset of $h$ failed nodes. The corresponding codes are labeled as $\cC_{h,k+1}^{(0)}$ and $\cC_{h,k+1},$ respectively.

Finally, in Section~\ref{sect:fg}, we present the main result of this paper---the construction for general values of both $h$ and $d$. In Section~\ref{sect:hd0} we construct an MDS code $\cC_{h,d}^{(0)}$ that supports
optimal repair of the first $h$ nodes, and in Section~\ref{sect:hd1} we use it as a building block to construct an $(n,k)$ MDS codes $\cC_{h,d}$ with the $(h,d)$-optimal repair property for general values of $h$ and $d$, $2\le h \le n-d\le r-1$.

The extension from repairing a fixed $h$-subset of nodes to any subset of cardinality $h$ relies on an idea that has already
appeared in the literature on regenerating codes \cite{Ye16,Goparaju17}, albeit in a somewhat veiled form. We isolate and illustrate this idea in 
Section~\ref{sect:hew}. Apart from revealing the structure behind our constructions, it also enables us to give a family
of $(n,k)$ {\em universal MSR codes} with the $(h,d)$-optimal repair property for all $1\le h\le n-d\le n-k$ simultaneously, i.e., these codes can optimally repair any number of failed nodes from any number of helper nodes. This construction forms a simple extension of the main results, and is given in a brief Section~\ref{sect:universal}.

Note that Sections~\ref{sect:bdblock}-\ref{sect:h} serve as preparation for Section~\ref{sect:fg}, and all the constructions in Sections~\ref{sect:bdblock}-\ref{sect:h} are special cases of the constructions in Section~\ref{sect:fg}. Even though the structure of the sections looks similar, each of the constructions adds new elements to the basic idea, and without the introductory sections it may be difficult to understand the intuition behind the code constructions in 
later parts of the paper. At the same time, we note that the codes in Sections~\ref{sect:fg} reduce to the codes in Section~\ref{sect:gend} and \ref{sect:h} upon appropriate adjustment of the parameters, such as taking $d=k+1$ or $h=2$, etc. (see Section~\ref{sect:connections} below for more details). 
The complete reduction scheme between the code families in this paper is as shown in Fig.~1, and the parameters of the codes are listed in Table~\ref{table:parameters}.

\begin{table}[ht]
\captionsetup{width=.8\linewidth,font=scriptsize}
\centering
\begin{tabular}{|l||cc|cc|}
\hline
&\multicolumn{2}{|c|}{Repairing the first $h$ nodes}&\multicolumn{2}{c|}{Repairing any $h$ nodes}\\
\hline
 \multicolumn{1}{|c||}{Values of $h=|\cF|,d=|\cR|$} &$|F|$ &$l$ &\hspace*{.1in}$|F|$ &$l$\\
 \hline
 Sec.~\ref{sect:bdblock}: $h=2,d=k+1$ &$n+2$ &3 &\hspace*{.1in}$2n$ &$3^{\binom n2}$\\
 Sec.~\ref{sect:gend}: $h=2,$ any $d$ &$n+2(s-1)$ &$s^2-1$ &\hspace*{.1in}$sn$ &$(s^2-1)^{\binom n2}$\\
 Sec.~\ref{sect:h}: any $h,$ $d=k+1$ &$n+h$ &$h+1$ &\hspace*{.1in}$2n$ &$(h+1)^{\binom n h}$\\
 Sec.~\ref{sect:fg}: any $h$, any $d$ &$n+h(s-1)$ &$(h+d-k)(s-1)^{h-1}$ &\hspace*{.1in}$sn$ &$((h+d-k)(s-1)^{h-1})^{\binom n h}$\\
 \hline
 \end{tabular}
 \caption{\noindent\hangindent .5in \hangafter=1 We list the parameters (field size, sub-packetization) of the codes constructed in this paper, where $s:=d+1-k$. In the first of the two pairs of columns the codes are constructed for optimal repair of the {\em first $h$ nodes only}, while the second pair gives the parameters of codes that can optimally repair {\em any} $h$ failed nodes.}\label{table:parameters}
\end{table}

\begin{center}
\begin{tikzpicture}
\draw node at (9.5,-1.5) [text width=0.3in, align=center](A) {$\cC_{2,k+1}^{(0)}$ {\scriptsize Sec.~\ref{sect:first}}};
\draw node at (7,0.5) [text width=0.3in, align=center](B) {$\cC_{2,d}^{(0)}$ {\scriptsize Sec.~\ref{sect:fd}}};
\draw node at (9.5,0.3) [text width=0.3in, align=center](C) {$\cC_{2,k+1}$ {\scriptsize Sec.~\ref{sect:warmup}}};
\draw node at (12,0.5) [text width=0.3in, align=center](D) {$\cC_{h,k+1}^{(0)}$ {\scriptsize Sec.~\ref{sect:gh}}};
\draw node at (7,2.5) [text width=0.3in, align=center](E) {$\cC_{2,d}$ {\scriptsize Sec.~\ref{sect:rb}}};
\draw node at (12,2.5) [text width=0.3in, align=center](F) {$\cC_{h,k+1}$ {\scriptsize Sec.~\ref{sect:lo}}};
\draw node at (9.5,2.3) [text width=0.3in, align=center](G) {$\cC_{h,d}^{(0)}$ {\scriptsize Sec.~\ref{sect:hd0}}};
\draw node at (9.5,4) [text width=0.3in, align=center](H) {$\cC_{h,d}$ {\scriptsize Sec.~\ref{sect:hd1}}};
\draw[->, >=stealth,line width=.2mm] (B) -- (A) node[draw=none,fill=none,font=\scriptsize,midway,below] {};
\draw[->, >=stealth,line width=.2mm] (C) -- (A) node[draw=none,fill=none,font=\scriptsize,near start,above] {};
\draw[->, >=stealth,line width=.2mm] (D) -- (A) node[draw=none,fill=none,font=\scriptsize,midway,below] {};
\draw[->, >=stealth,line width=.2mm] (E) -- (B) node[draw=none,fill=none,font=\scriptsize,midway,below] {};
\draw[->, >=stealth,line width=.2mm] (E) -- (C) node[draw=none,fill=none,font=\scriptsize,midway,above] {};
\draw[->, >=stealth,line width=.2mm] (F) -- (C) node[draw=none,fill=none,font=\scriptsize,midway,below] {};
\draw[->, >=stealth,line width=.2mm] (F) -- (D) node[draw=none,fill=none,font=\scriptsize,midway,below] {};
\draw[->, >=stealth,line width=.2mm] (G) -- (B) node[draw=none,fill=none,font=\scriptsize,midway,below] {};
\draw[->, >=stealth,line width=.2mm] (G) -- (D) node[draw=none,fill=none,font=\scriptsize,midway,below] {};
\draw[->, >=stealth,line width=.2mm] (H) -- (G) node[draw=none,fill=none,font=\scriptsize,midway,below] {};
\draw[->, >=stealth,line width=.2mm] (H) -- (E) node[draw=none,fill=none,font=\scriptsize,midway,below] {};
\draw[->, >=stealth,line width=.2mm] (H) -- (F) node[draw=none,fill=none,font=\scriptsize,midway,below] {};
\end{tikzpicture}

\noindent\begin{minipage}{.75\linewidth}{\footnotesize \noindent\hangindent=.35in\hangafter=1
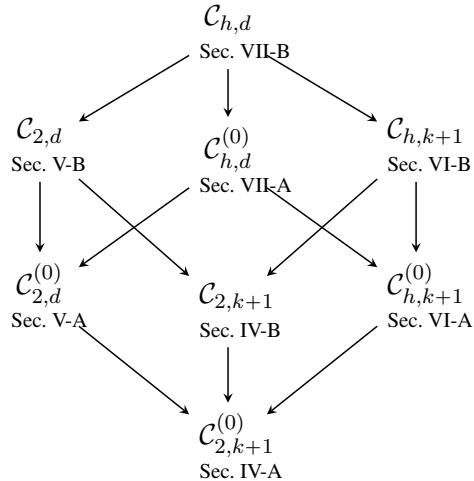
Fig.1: Relations between the code families constructed in the paper. Arrows point from more general
code families to their subfamilies. The superscript $^{(0)}$ indicates that the code supports optimal repair  of the first two (or the first $h$)
erasures only. \par}\end{minipage}
\end{center}

\subsection{Future directions}
\begin{enumerate}

\item In this paper we consider the problem of repairing multiple erasures for MDS codes, which correspond to the minimum storage regenerating (MSR) point on the trade-off curve between storage and repair bandwidth in the regenerating code literature \cite{Dimakis10,Elyasi16}. A natural future direction is to extend our results to the whole trade-off curve, starting with the minimum bandwidth regenerating (MBR) point.

\item The repair problem of Reed-Solomon (RS) codes has attracted significant attention recently \cite{Guruswami16,Dau17,Dau16,Chowdhury17,Bartan17,Ye16b,Tamo17RS,Ye17,Tamo18}. In particular, explicit RS code constructions with the $(h,d)$-optimal repair property under the centralized model were given in \cite{Ye17}. Can this result be extended to the cooperative model (and are two rounds enough)? Note that  cooperative repair of (full-length) RS codes was previously considered in \cite{Dau16}, which gave schemes for repairing 2 and 3 erasures with small repair bandwidth (since codes in \cite{Dau16} have small $l$, the repair bandwidth ends up being rather far away from the cut-set bound).

\item Let us consider the regime where we fix the number of parity nodes $r:=n-k$ and let $n$ grow.
The sub-packetization value of our MDS code construction with the $(h,d)$-optimal repair property scales as $\exp(\Theta(n^h))$ in this regime, which is much larger than its counterpart under the centralized model, where the sub-packetization value is $\exp(O(n))$ (see \cite{Ye16}). One possible reason is that since the cooperative model is more restrictive than the centralized model, the larger sub-packetization is the penalty we have to pay. The other possibility is that our construction can be improved in terms of the sub-packetization value. This raises an open question of either deriving a lower bound on sub-packetization for the cooperative model (cf. also Table~\ref{table:parameters}) or constructing codes with smaller sub-packetization.

\item Several families of codes under centralized repair also have the {\em optimal access} property, wherein the number
of field symbols accessed at the helper nodes equals the number of symbols downloaded for the purposes of repair \cite{Sasid16,Ye16a}.
Is it possible to design optimal-repair codes for the cooperative model that reduce or minimize the number of symbols accessed during the repair process?
\end{enumerate}

\section{Proof of \eqref{eq:cutset} and Theorem~\ref{thm:st}} \label{ap:st}

\remove{Although our proof below is written for two-round cooperative repair schemes, a simple modification allows us to apply the same proof to cooperative repair with any number of communication rounds, which means that Theorem~\ref{thm:st} holds for multiple-round cooperative repair; see Remark~\ref{rm:mr} for details.}

Let $\cC$ be an $(n,k,l)$ MDS code over $F$. Our goal is to prove that if \eqref{eq:cutset} holds with equality, then so does \eqref{eq:csce}. We will argue
by showing that inequality \eqref{eq:csce} implies \eqref{eq:cutset} and then observe that the equality in \eqref{eq:cutset} 
implies the same for \eqref{eq:csce}.  The first step of this
argument also yields a self-contained proof of the cooperative cut-set bound \eqref{eq:cutset}.

Recall that $h:=|\cF|$ and $d:=|\cR|$. To shorten the expressions, below we use the following notation
   $$
   D_i(\cR)= \sum_{j\in\cR}\dim_F(\big( f_{ij}(C_j) \big), \quad D_{i}(\cF)= \sum_{i'\in\cF\setminus\{i\}} \dim_F \big( f_{ii'}(\{f_{i'j}(C_j),j\in\cR \}) \big)
   $$
for the number of symbols of $F$ downloaded by $C_i\in \cF$ from the helper nodes (in the first round of repair) and
from the other failed nodes (in the second round of repair), respectively, where the functions $f_{i,\cdot}$ were introduced in 
Definition~\ref{def:op}. For a given node $C_i$ there are $d+h-1$ such functions, and therefore, in total there are $h(d+h-1)$
of them for any given subsets $\cF,\cR.$ Our goal is to show that 
%
\begin{equation}\label{eq:gl}
\sum_{i\in\cF}( D_i(\cR) 
+ D_{i}(\cF) )
\ge \frac{h(h+d-1)}{h+d-k}l.
\end{equation}

Our proof relies on the following simple observation: in the first round of the repair process, the data downloaded from the helper nodes by all the failed nodes is the following set of vectors:
\begin{equation}\label{eq:rm}
\{f_{ij}(C_j),i\in\cF,j\in\cR\}.
\end{equation}
After obtaining this set of vectors, the failed nodes can recover their values by performing additional information exchange
during the second round of repair. Recalling the centralized model, this means that all the information needed 
to collectively repair the failed nodes is contained in the set \eqref{eq:rm}. Therefore, on account of the
centralized version of the cut-set bound \eqref{eq:csce} we have
\begin{equation}\label{eq:pl1}
\sum_{i\in\cF} D_i(\cR) \ge \frac{hd}{h+d-k}l.
\end{equation}

To bound the second term on the left-hand side of \eqref{eq:gl}, we use the following basic fact about MDS code: for an $(n,k)$ MDS code, any subset of $k-1$ coordinates contains no information about any other coordinate of the code. Assume a uniform distribution
on the codewords $C=(C_1,\dots,C_n) \in \cC$ and (by a slight abuse of notation) use the same symbols $C_i, i=1,\dots,n$ for the associated random variables.
For any $i\in [n]$ (in particular, for any $i\in\cF$) and any subset $\cS\subseteq \cR$ of the helper nodes of size $|\cS|=k-1$, we have
  $$
H(C_i)=H(C_i|\{C_j,j\in\cS\})=l \log_2|F|,
  $$
where $H(X|Y)$ is the conditional entropy of $X$ given $Y$, measured in bits. Applying a deterministic function to $Y$ 
can only increase the conditional entropy, and therefore for any $\cS\subseteq\cR, |\cS|=k-1$ we have
  \begin{equation}\label{eq:k1}
H(C_i|\{f_{ij}(C_j),j\in\cS\})=l \log_2(|F|).
   \end{equation}
On the other hand, each $C_i,i\in\cF$ is uniquely determined by 
$\{f_{ij}(C_j),j\in\cR\}\cup\{f_{ii'}(\{f_{i'j}(C_j),j\in\cR \}):i'\in\cF\setminus\{i\}\}$, so
\begin{equation}\label{eq:k2}
H(C_i|\{f_{ij}(C_j),j\in\cR\}\cup\{f_{ii'}(\{f_{i'j}(C_j),j\in\cR \}):i'\in\cF\setminus\{i\}\})=0.
\end{equation}
Combining \eqref{eq:k1} and \eqref{eq:k2}, and using Lemma~\ref{lem:ax} below, we obtain that
\begin{align}\label{eq:H}
H \left( \{f_{ij}(C_j),j\in\cR\setminus\cS\}\cup\{f_{ii'}(\{f_{i'j}(C_j),j\in\cR \}):i'\in\cF\setminus\{i\}\} \right)
\ge l \log_2|F|.
\end{align}
Therefore, for any $i\in\cF$ and any $\cS\subseteq \cR, |\cS|=k-1$
\begin{align}
\sum_{j\in\cR\setminus\cS} \dim_F \big( f_{ij}(C_j) \big) 
+ \sum_{i'\in\cF\setminus\{i\}} \dim_F \big( f_{ii'}(\{f_{i'j}(C_j),j\in\cR \}) \big) \ge l \label{eq:js}
\end{align}
(the left-hand side on the above line is the entropy of the left-hand side of \eqref{eq:H} under the uniform distribution on 
its arguments. Since the entropy is maximized for the uniform distribution, \eqref{eq:js} is implied by \eqref{eq:H}. Note also
the switching of the base of logarithms from 2 to $|F|$.).

 Let us sum \eqref{eq:js} over all subsets $\cS\subseteq\cR$ of size $|\cS|=k-1$.
Only the first term on the left-hand side depends on $\cS$, and for every $j\in\cR$, the term $\dim_F \big( f_{ij}(C_j) \big)$ appears for $\binom{d-1}{k-1}$ different choices of $\cS.$ Thus we have
$$
\binom{d-1}{k-1} D_i(\cR) 
+ \binom{d}{k-1} D_i(\cF) \ge \binom{d}{k-1} l, \quad i\in \cF.
$$
Dividing both sides by $\binom{d}{k-1},$ we obtain that for every $i\in\cF$,
$$
\frac{d-k+1}{d} D_i(\cR) 
+  D_i(\cF) \ge l.
$$
Let us sum these inequalities on all $i\in\cF$. We obtain
\begin{equation}\label{eq:pl2}
\frac{d-k+1}{d} \sum_{i\in\cF}  D_i(\cR) 
+  \sum_{i\in\cF} D_i(\cF) \ge hl.
\end{equation}
Multiplying \eqref{eq:pl1} on both sides by $\frac{k-1}{d}$ and then adding it to \eqref{eq:pl2}, we obtain the desired inequality \eqref{eq:gl}.
This completes the proof of \eqref{eq:cutset}.

We are left to prove the claim that for a given code $\cC$, \eqref{eq:of} implies \eqref{eq:xv}. Assuming $\eqref{eq:of},$ 
we observe that there is a choice of the functions $\{\{f_{ij}, j\in \cR\}, \{f_{ii'}, i'\in \cF\backslash\{i\}\}: i\in \cF\}$ such that \eqref{eq:gl} holds with equality. This means that \eqref{eq:pl2} and all the inequalities preceding it in the proof, including
 \eqref{eq:pl1}, hold with equality, but equality in \eqref{eq:pl1} means that \eqref{eq:xv} holds true.


\begin{lemma}\label{lem:ax} Let $X,Y,Z$ be arbitrary discrete random variables such that $H(X|YZ)=0,$ then $H(Z)\ge H(X|Y).$
\end{lemma}
\begin{IEEEproof}  By the assumption we have $H(XYZ)=H(YZ)$. Therefore,
  \begin{align*}
H(Z)\ge H(Z|Y)&= H(YZ)-H(Y)\\& = H(XYZ)-H(Y)\\&  \ge H(XY)-H(Y) \\& = H(X|Y).
\end{align*}
\end{IEEEproof}

It remains to justify the final claim of the theorem, namely that it holds for the general case of $T\ge 2$ 
communication rounds. Indeed the proof given above can be easily modified
to cover the general situation.
To explain this, let us assume that the repair process is divided into $T$ rounds for some finite integer $T$. 
In this case, for $i\in\cF$ and $j\in\cR$, we view $f_{ij}(C_j)$ as all the data downloaded by the failed node 
$C_i$ from the helper node $C_j$ in all $T$ rounds of communication.
For $i,i'\in \cF,i\neq i'$, we view $f_{ii'}(\{f_{i'j}(C_j),j\in\cR \})$ as all the data downloaded by the failed node $C_i$ from 
another failed node $C_{i'}$ in all $T$ rounds of communication\footnote{Observe
 that the notation $f_{ii'}(\{f_{i'j}(C_j),j\in\cR \})$ 
is not accurate for multiple-round repair because $f_{ii'}$ can also depend on the data $f_{i'j},j\in\cF\setminus\{i'\}$ downloaded in previous round(s). At the same time, this issue 
does not affect our argument, so we prefer to keep the already established notation.}.
It is easy to check that under this point of view, our proof applies directly to a $T$-round repair process for any integer $T$.

\section{A technical lemma}\label{sect:sl}
In this section we prove a technical lemma which will be frequently used throughout the paper. 
Let $C\in \cC$ be a codeword of an $(n,k=n-r,l)$ MDS array code $\cC$. We write $C$ as $(C_1,C_2,\dots,C_n)$,
where $C_i=(c_{i,0},c_{i,1},\dots,c_{i,l-1})^T \in F^l$ is the $i$th coordinate of $C$.
\begin{lemma}\label{lem:tch}
Let $n,k,d$ be positive integers such that $k\le d \le n-1$. Let $r:=n-k$ and let $s:=d+1-k$.
Let $F$ be a finite field with cardinality $|F|\ge n+s-1$. Let $\lambda_{1,0},\lambda_{1,1},\dots,\lambda_{1,s-1}, \lambda_2,\lambda_3,\dots,\lambda_n$ be $n+s-1$ distinct elements of $F$.
Define an $(n,k,s)$ MDS array code $\cC$ over the field $F$ by the following $rs$ parity check equations:
\begin{equation}\label{eq:org}
\lambda_{1,u}^t c_{1,u} + \sum_{i=2}^n \lambda_i^t c_{i,u} = 0, \quad
u=0,1,\dots,s-1, \quad t=0,1,\dots,r-1.
\end{equation}
Let $\mu_i:= \sum_{u=0}^{s-1} c_{i,u}$ for all $i\in[n]$. Then for any subset $\cR\subseteq \{2,3,\dots,n\}$ with cardinality $|\cR|=d$, the values $\{c_{1,0},c_{1,1},\dots,c_{1,s-1},\mu_2,\mu_3,\dots,\mu_n\}$ can be calculated from $\{\mu_i:i\in\cR\}$.
\end{lemma}

\begin{IEEEproof}\!\footnote{This proof draws on the ideas in \cite[Theorem 7]{Ye16}.}
 Summing \eqref{eq:org} over $u\in\{0,1,\dots,s-1\}$, we obtain
   $$
\sum_{u=0}^{s-1} \lambda_{1,u}^t c_{1,u} + \sum_{i=2}^n \lambda_i^t \mu_i = 0,
\quad    t=0,1,\dots,r-1.
   $$
Writing these $r$ equations in matrix form, we obtain the following equality:
\begin{equation}\label{eq:ml}
 \left[\begin{array}{ccccccccc}
1 & 1 & \dots & 1 & 1 & 1 & 1 & \dots & 1\\
\lambda_{1,0} & \lambda_{1,1} & \dots & \lambda_{1,s-1} & \lambda_2 & \lambda_3 & \lambda_4 & \dots & \lambda_n \\
\lambda_{1,0}^2 & \lambda_{1,1}^2 & \dots & \lambda_{1,s-1}^2 & \lambda_2^2 & \lambda_3^2 & \lambda_4^2 & \dots & \lambda_n^2 \\
\vdots & \vdots & \vdots & \vdots & \vdots & \vdots & \vdots & \vdots & \vdots \\
\lambda_{1,0}^{r-1} & \lambda_{1,1}^{r-1} & \dots & \lambda_{1,s-1}^{r-1} & \lambda_2^{r-1} & \lambda_3^{r-1} & \lambda_4^{r-1} & \dots & \lambda_n^{r-1} \end{array}\right]
\left[\begin{array}{c} c_{1,0} \\ c_{1,1} \\ \vdots \\ c_{1,s-1}\\
\mu_2\\ \mu_3\\ \mu_4\\ \vdots \\ \mu_n\end{array}\right] = 0.
\end{equation}
Since $\lambda_{1,0}, \lambda_{1,1}, \dots, \lambda_{1,s-1}, \lambda_2, \lambda_3, \lambda_4, \dots, \lambda_n$ are all distinct, the vector 
$(c_{1,0},c_{1,1},\dots,c_{1,s-1},\mu_2,\mu_3,\dots,\linebreak[4] \mu_n)$ is a codeword in an $(n+s-1,n+s-1-r=d)$ generalized Reed-Solomon code. 
Therefore, for any $\cR\subseteq \{2,3,\dots,n\}, |\cR|=d,$ 
 the values $\{c_{1,0},c_{1,1},\dots,c_{1,s-1},\mu_2,\mu_3,\dots,\mu_n\}$ can be calculated from $\{\mu_i:i\in\cR\}$. This completes the proof of the lemma.
\end{IEEEproof}

\section{Cooperative $(2,k+1)$-optimal codes}\label{sect:bdblock}
\subsection{Repairing the first two nodes from any $k+1$ helper nodes}\label{sect:first}
Let $F$ be a finite field. For any $k<n\le |F|-2$ we present a construction of $(n,k,3)$ MDS array codes $\cC=\cC_{2,k+1}^{(0)}$ over $F$ that support
optimal repair of the first two nodes. Specifically, when the first two nodes of $\cC$ fail, the repair of each failed node can be accomplished by connecting to {\em any} $k+1$ helper nodes and downloading a total of $k+2$ symbols of $F$ 
from these helper nodes as well as from the other failed node, achieving the optimal repair bandwidth according to the cut-set bound \eqref{eq:cutset}.

For $i=1,2,\dots,n$, we write the $i$th node of $\cC$ as $C_i=(c_{i,0},c_{i,1}, c_{i,2})^T\in F^3$, which is a column vector of dimension $3$ over $F$.
Let $\lambda_{1,0},\lambda_{1,1},\lambda_{2,0},\lambda_{2,1},\lambda_3,\lambda_4,\dots,\lambda_n$ be $n+2$ distinct elements of the field $F$.
The code $\cC$ is defined by the following $3$ sets of parity check equations: 
\begin{align}
\lambda_{1,0}^t c_{1,0} + \lambda_{2,0}^t c_{2,0} + \sum_{i=3}^n \lambda_i^t c_{i,0} = 0, 
\quad t=0,1,\dots,r-1, \label{eq:c11} \\
\lambda_{1,1}^t c_{1,1} + \lambda_{2,0}^t c_{2,1} + \sum_{i=3}^n \lambda_i^t c_{i,1} = 0,     
\quad t=0,1,\dots,r-1, \label{eq:c12} \\
\lambda_{1,0}^t c_{1,2} + \lambda_{2,1}^t c_{2,2} + \sum_{i=3}^n \lambda_i^t c_{i,2} = 0, 
\quad t=0,1,\dots,r-1. \label{eq:c13}
\end{align}
For each $a=0,1,2$ the set of vectors $\{(c_{1,a},c_{2,a},\dots,c_{n,a})\}$ obviously forms an $(n,k=n-r)$ MDS code, and so  $\cC$ is indeed
an $(n,k,3)$ MDS array code.

The following lemma suggests a description of the repair scheme for the first two nodes using the bandwidth that meets the cut-set bound \eqref{eq:cutset} with equality.
\begin{lemma}\label{lem:bb} For $i=1,\dots,n$ let 
$$
\mu_{i,1}:=c_{i,0}+c_{i,1}, \quad
\mu_{i,2}:=c_{i,0}+c_{i,2}.
$$
For any set of helper nodes $\cR\subseteq \{3,4,\dots,n\},|\cR|=k+1$, the values of $c_{1,0},c_{1,1},$ and $\mu_{2,1}$ are uniquely determined by $\{\mu_{i,1}:i\in \cR\}$.
Similarly, the values of $c_{2,0},c_{2,2},$ and $\mu_{1,2}$ are uniquely determined by $\{\mu_{i,2}:i\in \cR\}$.
\end{lemma}
\begin{IEEEproof}
This lemma follows immediately from Lemma~\ref{lem:tch}. Indeed, take $d=k+1$ and $s=2$, then there are only two groups of 
equations in \eqref{eq:org}, namely those for $u=0,1.$
To prove the first statement of Lemma~\ref{lem:bb}, consider the equations in \eqref{eq:c11} and \eqref{eq:c12}. These two sets of equations have the same structure as the equations in \eqref{eq:org}: namely,
only the coefficients of $c_{1,u}$ vary with $u$ while the coefficients of $c_{i,u}$ are 
independent of the value of $u$ for all $i\in\{2,3,\dots,n\}$. Therefore Lemma~\ref{lem:tch} applies directly, and we obtain the
claimed fact about $c_{1,0},c_{1,1}$ and $\mu_{2,1}.$

Similarly, to prove the second statement, consider the equations in \eqref{eq:c11} and \eqref{eq:c13}. These two sets of equations also have the same structure as the equations in \eqref{eq:org}: namely,
only the coefficients of $c_{2,u}$ vary with $u$ while the coefficients of $c_{i,u}$ are independent of the value of $u$ for all $i\in[n]\setminus\{2\}$.
\end{IEEEproof}
This lemma implies 
that the first two nodes of $\cC$ can be repaired with optimal bandwidth.
As already mentioned, the repair process is divided into two rounds. 
In the first round, the node $C_j,j=1,2$ downloads $k+1$ symbols 
$\mu_{ij}$ from the helper nodes $C_i,i\in \cR$.
According to Lemma~\ref{lem:bb}, after the first round, $C_1$ knows the values of
$c_{1,0},c_{1,1}$ and $c_{2,0}+c_{2,1}$, and $C_2$ knows the values of $c_{2,0},c_{2,2}$ and $c_{1,0}+c_{1,2}$.
In the second round, $C_1$ downloads the sum $c_{1,0}+c_{1,2}$ from $C_2$, and $C_2$ downloads the sum $c_{2,0}+c_{2,1}$ from $C_1$. Clearly, after the second round, both $C_1$ and $C_2$ can recover all their coordinates. Moreover, in the whole repair process, $C_1$ only downloads one symbol of $F$ from each of the nodes $C_i,i\in \cR\cup\{2\}$, and $C_2$ only downloads one symbol of $F$ from each of the nodes $C_i,i\in \cR\cup\{1\}$.
Therefore the total repair bandwidth is $2(k+1)+2$, meeting the cut-set bound \eqref{eq:cutset} with equality.

\subsection{Repairing any two erasures from any $k+1$ helper nodes}\label{sect:warmup}
Here we develop the idea in the previous section to construct explicit MDS array codes with the $(2,k+1)$-optimal repair property. 
More specifically, given any $n\ge k+3$ and a finite field $F,|F|\ge 2n$, we present an $(n,k,l=3^{m})$ MDS array code $\cC=\cC_{2,k+1}$ over 
$F,$ where $m=\binom n2.$
 When {\em any} two nodes of $\cC$ fail, the repair of each failed node can be accomplished by connecting to {\em any} $k+1$ helper nodes and downloading $(k+2)3^{m-1}$ symbols of $F$ in total from these helper nodes as well as from the other failed node.
Clearly, the repair bandwidth meets the cut-set bound \eqref{eq:cutset} with equality.

We will define $\cC$ by its parity-check equations, and we begin with some notation. Let $\{\lambda_{i,j}\}_{i\in[n],j\in\{0,1\}}$ be $2n$ distinct elements of the field $F$.
Let $g$ be a bijection between the set of pairs $\{(i_1,i_2):1\le i_1<i_2\le n\}$ and the set $\{1,2,\dots,m\}$. For concreteness, let 
  \begin{equation}\label{eq:bn}
  g: (i_1,i_2)\mapsto\binom{i_2-1}{2}+i_1
  \end{equation}
($g$ partitions the set $[m]$ into
segments of length $(i_2-1)$, where $i_2=2,3,\dots,n$).
Given an integer $a\in\{0,1,\dots,l-1\}$, let $(a_m,a_{m-1},\dots, a_1)$ be the digits of its ternary expansion, i.e., 
$a=\sum_{j=0}^{m-1}a_{j+1}3^j.$ 
Define the following function 
\begin{equation}\label{eq:deff}
\begin{aligned}
f:&\,[n]\times\{0,1,\dots,l-1\}\to\{0,1\}\\
&(i,a)\mapsto\Big(\sum_{j=1}^{i-1} \mathbbm{1}\{a_{g(j,i)}=2\}+ \sum_{j=i+1}^n
\mathbbm{1}\{a_{g(i,j)}=1\} \Big) \Mod 2,
\end{aligned}
\end{equation}
where $\mathbbm{1}$ is the indicator function. We note that $f$ computes the parity of the count of 1's and 2's in a certain subset of the digits of $a.$ This subset is formed of all the digits with indices in the set $\{g(1,i),\dots,g(i-1,i),g(i,i+1),\dots,g(i,n)\}$. To give an example, let $n=6,$ then $m=15$, and the function $g$ maps from $\{(i_1,i_2):1\le i_1<i_2\le 6\}$ to $\{1,2,\dots,15\}$. Let $i=2$ and let $0\le a\le 3^{15}-1= 14348906$ be an integer. The function $f$ isolates the digits $a_u$ in the ternary expansions of $a$ such that $u\in\{g(\cdot,2),g(2,\cdot)\},$
i.e., $u\in\{g(1,2),g(2,3),g(2,4),g(2,5),g(2,6)\}=\{1,3,5,8,12\}.$ The value of the function $f(2,a)$ equals the parity of
$\mathbbm{1}\{a_1=2\}+\mathbbm{1}\{a_3=1\}+\mathbbm{1}\{a_5=1\}+\mathbbm{1}\{a_8=1\}+\mathbbm{1}\{a_{12}=1\}.$

\begin{definition}\label{def:anytwo}
The code $\cC=\cC_{2,k+1}$ is defined by the following $rl$ parity check equations:
$$
\sum_{i=1}^n \lambda_{i,f(i,a)}^t c_{i,a} = 0,\;
t=0,1,\dots,r-1, a=0,1,\dots,l-1.
$$
\end{definition}
For all $a=0,1,\dots,l-1$, the set of vectors 
$\{(c_{1,a},c_{2,a},\dots,c_{n,a})\}$ forms an $(n,k)$ MDS code, 
so $\cC$ is indeed an $(n,k,l)$ MDS array code.

Next we show that $\cC$ has optimal repair bandwidth for repairing any two failed nodes from any $k+1$ helper nodes.
Let $C_{i_1}$ and $C_{i_2}, i_1<i_2$ be the failed nodes. 
First let us introduce some notation to describe the repair scheme.
For $a=0,1,\dots,l-1$, $j\in[m],$ and  $u=0,1,2,$ let 
   $$
   a(j,u):=(a_m,\dots,a_{j+1},u,a_{j-1},\dots,a_1).
   $$
For $a=0,1,\dots,l-1$ and $i\in[n]$, let
  \begin{align*}
\mu_{i,1}^{(a)}&:=c_{i,a(g_{12},0)}+c_{i,a(g_{12},1)}, \\
\mu_{i,2}^{(a)}&:=c_{i,a(g_{12},0)}+c_{i,a(g_{12},2)},
  \end{align*}
  where for brevity we write $g_{12}$ instead of $g(i_1,i_2).$

The fol\-low\-ing lemma, which develops the ideas in Lemma~\ref{lem:bb}, accounts for the $(2,k+1)$ optimal repair property of the code $\cC.$ 
  \begin{lemma}\label{lem:cv} Let $C_{i_1}$ and $C_{i_2},$ $i_1<i_2$ be the failed nodes.
For any set of helper nodes $\cR\subseteq [n]\setminus\{i_1,i_2\}, |\cR|=k+1$ and any $a\in\{0,1,\dots,l-1\}$, the values 
$
c_{i_1,a(g_{12},0)},c_{i_1,a(g_{12},1)},\mu_{i_2,1}^{(a)}
$
 are uniquely determined by the set of values $\{\mu_{i,1}^{(a)}:i\in \cR\}$.
Similarly, the values  
$
c_{i_2,a(g_{12},0)},c_{i_2,a(g_{12},2)},\mu_{i_1,2}^{(a)}
$
 are uniquely determined by the set of values $\{\mu_{i,2}^{(a)}:i\in \cR\}$.
   \end{lemma}
\begin{IEEEproof} Recall that $a=0,1,\dots,l-1$ numbers the coordinates of the node, or the rows in the codeword array. For a fixed value of $a$, 
the parity check equations corresponding to the rows $a(g_{12},0),a(g_{12},1),a(g_{12},2)$ are as follows:
\begin{equation}\label{eq:cis}
\sum_{i=1}^n \lambda_{i,f(i,a(g_{12},u))}^t c_{i,a(g_{12},u)} = 0
, \quad t=0,1,2,\dots,r-1, \quad u=0,1,2.
\end{equation}
According to definition of the function $f$ in \eqref{eq:deff} and the remarks made after it, 
we have
  \begin{align*}
  f(i,a(g_{12},0)) &= f(i,a(g_{12},1)) = f(i,a(g_{12},2)), \quad i\in[n]\setminus\{i_1,i_2\}\\
  f(i_1,a(g_{12},0)) &= f(i_1,a(g_{12},2)) \neq f(i_1,a(g_{12},1)), \\
f(i_2,a(g_{12},0)) &= f(i_2,a(g_{12},1)) \neq f(i_2,a(g_{12},2)).
  \end{align*}
This implies that for $i\in[n]\setminus\{i_1,i_2\}$ the following notation is well defined:
\begin{equation}\label{eq:sh1}
\lambda_i:=\lambda_{i,f(i,a(g_{12},0))} = \lambda_{i,f(i,a(g_{12},1))}
=\lambda_{i,f(i,a(g_{12},2))}.
\end{equation}
Note that $\lambda_i$ depends on the value of $a$, though we omit this dependence from the notation.
Further, let
\begin{equation}\label{eq:sh2}
\begin{aligned}
\lambda_{i_1,0}'&:= \lambda_{i_1,f(i_1,a(g_{12},0))} = \lambda_{i_1,f(i_1,a(g_{12},2))},\\
\lambda_{i_1,1}'&:= \lambda_{i_1,f(i_1,a(g_{12},1))}, \\
\lambda_{i_2,0}'&:= \lambda_{i_2,f(i_2,a(g_{12},0))} = \lambda_{i_2,f(i_2,a(g_{12},1))},\\
\lambda_{i_2,1}'&:= \lambda_{i_2,f(i_2,a(g_{12},2))}.
\end{aligned}
\end{equation}
Notice that
\begin{gather*}
\lambda_{i_1,0}'\neq \lambda_{i_1,1}',
\lambda_{i_2,0}'\neq \lambda_{i_2,1}'\\
\{\lambda_{i_1,0}', \lambda_{i_1,1}'\}=\{\lambda_{i_1,0}, \lambda_{i_1,1}\}\\
\{\lambda_{i_2,0}', \lambda_{i_2,1}'\}=\{\lambda_{i_2,0}, \lambda_{i_2,1}\}\\
\lambda_i\in\{\lambda_{i,0},\lambda_{i,1}\},\; i\in[n]\setminus\{i_1,i_2\}.
\end{gather*}
Therefore $\lambda_{i_1,0}',\lambda_{i_1,1}',\lambda_{i_2,0}',\lambda_{i_2,1}',
\lambda_i,i\in[n]\setminus\{i_1,i_2\}$ are all distinct.
Using the notation defined in \eqref{eq:sh1}-\eqref{eq:sh2}, we can write \eqref{eq:cis} as
\begin{align*}
(\lambda_{i_1,0}')^t c_{i_1,a(g_{12},0)}
+ (\lambda_{i_2,0}')^t c_{i_2,a(g_{12},0)}
+\sum_{i\in[n]\setminus\{i_1,i_2\}} \lambda_i^t c_{i,a(g_{12},0)} &= 0, \\
(\lambda_{i_1,1}')^t c_{i_1,a(g_{12},1)}
+ (\lambda_{i_2,0}')^t c_{i_2,a(g_{12},1)}
+\sum_{i\in[n]\setminus\{i_1,i_2\}} \lambda_i^t c_{i,a(g_{12},1)} &= 0, \\
(\lambda_{i_1,0}')^t c_{i_1,a(g_{12},2)}
+ (\lambda_{i_2,1}')^t c_{i_2,a(g_{12},2)}
+\sum_{i\in[n]\setminus\{i_1,i_2\}} \lambda_i^t c_{i,a(g_{12},2)} &= 0,\\
t&=0,1,2,\dots,r-1.
\end{align*}
Now notice that up to a notational change, these equations have the same form as equations \eqref{eq:c11}-\eqref{eq:c13}. 
Therefore, the proof of Lemma~\ref{lem:bb} applies directly, completing the proof.
\end{IEEEproof}

This lemma implies that the nodes $C_{i_1}$ and $C_{i_2}$ can be repaired with optimal bandwidth.
To see this, we partition the coordinates of a node into $l/3$ groups of
size $3$ where each group is formed of the coordinates with indices $a(g_{12},0),a(g_{12},1),a(g_{12},2)$ for a given $a$. By Lemma~\ref{lem:cv} above we know that each group can be repaired with optimal bandwidth, so the entire contents of the failed nodes can also be 
 optimally recovered.

A more detailed description of the repair process is as follows.
 In the first round of the repair process, $C_{i_1}$ downloads the values in the set 
$\{\mu_{i,1}^{(a)}:a_{g_{12}}=0\}$ and  $C_{i_2}$ downloads the values $\{\mu_{i,2}^{(a)}:a_{g_{12}}=0\}$ from each helper node $C_i,i\in \cR$.
This enables $C_{i_1}$ to find the values
   $$
   \{c_{i_1,a}:a_{g_{12}}=0\}\cup\{c_{i_1,a(g_{12},1)}:a_{g_{12}}=0\}\cup\{\mu_{i_2,1}^{(a)}:a_{g_{12}}=0\}.
   $$
Similarly, $C_{i_2}$ is able to find the values 
   $$
   \{c_{i_2,a}:a_{g_{12}}=0\}\cup\{c_{i_2,a(g_{12},2)}:a_{g_{12}}=0\}\cup\{\mu_{i_1,2}^{(a)}:a_{g_{12}}=0\}.
   $$
In the second round, $C_{i_1}$ downloads $\{\mu_{i_1,2}^{(a)}:a_{g_{12}}=0\}$ from $C_{i_2}$, and $C_{i_2}$ downloads $\{\mu_{i_2,1}^{(a)}:a_{g_{12}}=0\}$ from $C_{i_1}$. After the second round, $C_{i_1}$ knows the values of all the elements in the set
  \begin{equation*}
\{c_{i_1,a(g_{12},u)}:a_{g_{12}}=0,u\in\{0,1,2\}\}
=\{c_{i_1,a}:a\in\{0,1,2,\dots,l-1\}\},
  \end{equation*}
and $C_{i_2}$ knows the values of all the elements in the set
  \begin{equation*}
\{c_{i_2,a(g_{12},u)}:a_{g_{12}}=0,u\in\{0,1,2\}\}
=\{c_{i_2,a}:a\in\{0,1,2,\dots,l-1\}\},
  \end{equation*}
i.e., both $C_{i_1}$ and $C_{i_2}$ can recover all their coordinates. Moreover, in the whole repair process, $C_{i_1}$ downloads $l/3$ symbols of $F$ from each of the nodes $C_i,i\in \cR\cup\{i_2\}$, and $C_{i_2}$ downloads $l/3$ symbols of $F$ from each of the nodes $C_i,i\in \cR\cup\{i_1\}$.
Therefore the total repair bandwidth is $2(k+2)l/3$, meeting the cut-set bound \eqref{eq:cutset} with equality.

\section{Cooperative $(2,d)$-optimal codes for general $d$}\label{sect:gend}
\subsection{Optimal repair of the first two nodes}\label{sect:fd}

In this section we present an explicit MDS array code that can optimally repair the first two nodes from any $d$ helper nodes for general values of $d$. Let $n,k,d$ be such that $k+1\le d \le n-2$, let $s:=d+1-k,$ and let $F$ be a finite field of
size at least $n-2+2s.$ We will construct an $(n,k,s^2-1)$ MDS array code $\cC=\cC_{2,d}^{(0)}$ over the field $F$ that has the following
property. When the first two nodes of $\cC$ fail, the repair of each of them can be accomplished by connecting to {\em any} $d$ 
surviving (helper) nodes and downloading $(s-1)(d+1)$ symbols of $F$ in total from these helper nodes as well as from
the other failed node. Clearly, the amount of downloaded data meets the cut-set bound \eqref{eq:cutset} with equality.

Let $\lambda_{1,0},\lambda_{1,1},\dots,\lambda_{1,s-1},
\lambda_{2,0},\lambda_{2,1},\dots,\lambda_{2,s-1},
\lambda_3,\lambda_4,\dots,\lambda_n$ be $n-2+2s$ distinct elements of the field $F$.
Given  an integer $a, 0\le a\le s^2-2,$ let $b_1(a),b_2(a)$ be the digits of its expansion to the base $s$:
  \begin{equation}  \label{eq:defb}
  a=(b_2(a),b_1(a)).
  \end{equation}
The code $\cC=\cC_{2,d}^{(0)}$ is defined by the following $r(s^2-1)$ parity check equations.
\begin{align}\label{eq:asj}
\lambda_{1,b_1(a)}^t c_{1,a} + \lambda_{2,b_2(a)}^t c_{2,a}
&+\sum_{i=3}^n \lambda_i^t c_{i,a} = 0.\\
&t=0,1,\dots,r-1 ,\; a=0,1,2,\dots,s^2-2.\nonumber
\end{align}
Clearly, for a given $a$ the set of vectors $\{(c_{1,a},c_{2,a},\dots,c_{n,a})\}$ that satisfy the system \eqref{eq:asj} forms an MDS code of length $n$ and dimension $k$.
Therefore $\cC$ is indeed an $(n,k,s^2-1)$ MDS array code.
Note that for $d=k+1$, the code $\cC$ defined by \eqref{eq:asj} is the same as the code defined by \eqref{eq:c11}-\eqref{eq:c13} in Section~\ref{sect:bdblock}.

For every $i\in[n]$ define the following elements of $F$:
\begin{align*}
\mu_{i,1}^{(v_2)}:=\sum_{v_1=0}^{s-1}c_{i,sv_2+v_1} , \quad v_2 \in\{0,1,\dots,s-2\}; \\
\mu_{i,2}^{(v_1)}:=\sum_{v_2=0}^{s-1}c_{i,sv_2+v_1} , \quad v_1 \in\{0,1,\dots,s-2\}.
\end{align*}
Similarly to the previous sections, we have the following lemma:
\begin{lemma}\label{lem:jj} Suppose that the failed nodes are $C_1,C_2$ and let
$\cR\subseteq \{3,4,\dots,n\},|\cR|=d$ be a set of $d$ helper nodes.
For any $v_2\in\{0,1,\dots,s-2\}$, 
the values $\{c_{1,sv_2+v_1}, v_1=0,1,\dots,s-1\}$ and $\mu_{2,1}^{(v_2)}$
 are uniquely 
determined by the set of values $\{\mu_{i,1}^{(v_2)}:i\in \cR\}$.
Similarly, for any $v_1\in\{0,1,\dots,s-2\}$, the values $\{c_{2,sv_2+v_1},v_2=0,1,\dots,s-1\}$ and 
$\mu_{1,2}^{(v_1)}$
are uniquely determined by the set of values $\{\mu_{i,2}^{(v_1)}:i\in \cR\}$.
\end{lemma}
\begin{IEEEproof}
We again use Lemma~\ref{lem:tch} to prove this lemma.
To prove the first statement, we use definition \eqref{eq:asj} to write out the parity-check equations that correspond to $a=sv_2,sv_2+1,\dots,sv_2+s-1$ for a fixed $v_2\in\{0,1,\dots,s-2\}$:
\begin{align*}
\lambda_{1,v_1}^t c_{1,sv_2+v_1} + \lambda_{2,v_2}^t c_{2,sv_2+v_1}
  &+\sum_{i=3}^n \lambda_i^t c_{i,sv_2+v_1} = 0,\\
&t=0,1,\dots,r-1,\; v_1=0,1,\dots,s-1.
\end{align*}
These equations have the same structure as the equations in \eqref{eq:org}: $v_1$ here plays the role of $u$ in \eqref{eq:org}.
Only the coefficients of $c_{1,sv_2+v_1}$ vary with the value of $v_1$ while the coefficients of $c_{i,sv_2+v_1}$ are independent of the value of $v_1$ for all $i\in[n]\setminus\{1\}$. Therefore the proof of Lemma~\ref{lem:tch} can be directly applied here.

To prove the second statement, we use definition \eqref{eq:asj} to write out the parity-check equations that correspond to $a=v_1,v_2+v_1, 2v_2+v_1,\dots,(s-1)v_2+v_1$ for a fixed $v_1\in \{0,1,\dots,s-2\}$:
\begin{align*}
\lambda_{1,v_1}^t c_{1,sv_2+v_1} + \lambda_{2,v_2}^t c_{2,sv_2+v_1}
  &+\sum_{i=3}^n \lambda_i^t c_{i,sv_2+v_1} = 0,\\
&t=0,1,\dots,r-1,\; v_2=0,1,\dots,s-1.
\end{align*}
These equations have the same structure as the equations in \eqref{eq:org}: $v_2$ here plays the role of $u$ in \eqref{eq:org}.
Only the coefficients of $c_{2,sv_2+v_1}$ vary with the value of $v_2$ while the coefficients of $c_{i,sv_2+v_1}$ are independent of the value of $v_2$ for all $i\in[n]\setminus\{2\}$. Therefore the proof of Lemma~\ref{lem:tch} can be directly applied here.
\end{IEEEproof}

Let us show that this lemma implies that the first two nodes of $\cC$ can be repaired with optimal bandwidth.
 In the first round, the first node $C_1$ downloads the values $\{\mu_{i,1}^{(v_2)}, v_2=0,1,\dots,s-2\}$
  from each helper node $C_i,i\in \cR$, and the second node $C_2$ downloads $\{\mu_{i,2}^{(v_1)}, v_1=0,1,\dots,s-2\}$ 
  from each helper node $C_i,i\in \cR$. 
From Lemma~\ref{lem:jj} we conclude that after the first round, $C_1$ knows the values 
     \begin{gather*}
     c_{1,sv_2+v_1},\; v_2=0,1,\dots,s-2,v_1=0,1,\dots,s-1\\  
   \text{and~}  \mu_{2,1}^{(v_2)},\;v_2=0,1,\dots,s-2.
     \end{gather*}
     In the same way, $C_2$ knows the values  
   \begin{gather*}
   c_{2,sv_2+v_1},\;v_1=0,1,\dots,s-2,v_2=0,1,\dots,s-1\\ 
   \mu_{1,2}^{(v_1)},\; v_1=0,1,\dots,s-2.
   \end{gather*}
In the second round, $C_1$ downloads the sums $\mu_{1,2}^{(v_1)}, v_1=0,1,\dots,s-2$  from $C_2$, and $C_2$ downloads the sums $\mu_{2,1}^{(v_2)}, v_2=0,1,\dots,s-2$
 from $C_1$. It is easy to verify that after the second round, both $C_1$ and $C_2$ can recover all of their coordinates. Moreover, over the course of the entire repair process, 
$C_1$ downloads $(s-1)$ symbols of $F$ from each of the nodes $C_i,i\in \cR\cup\{2\}$, and $C_2$ downloads $(s-1)$ symbols of $F$ from each of the nodes $C_i,i\in \cR\cup\{1\}$.
Therefore the total repair bandwidth is $2(s-1)(d+1)$, meeting the cut-set bound \eqref{eq:cutset} with equality.

\subsection{Optimal repair of any two erasures}\label{sect:rb}
In this section we present a construction of MDS array codes with the $(2,d)$-optimal repair property, relying on the ideas of
the previous section. Let $n,k,d$ be such that $k+1\le d \le n-2,$ let $s:=d+1-k$ and let $F$ be a finite field such that $|F|\ge sn.$
We present an $(n,k,l=(s^2-1)^m)$ MDS array code $\cC=\cC_{2,d}$ over the field $F$, where $m:=\binom{n}{2}$. 
When {\em any} two nodes of $\cC$ fail, the repair of each failed node can be accomplished by connecting to 
{\em any} $d$ helper nodes and downloading $(d+1)l/(s+1)$ symbols of $F$ in total from these helper nodes as well as from
the other failed node. Clearly, the repair bandwidth meets the cut-set bound \eqref{eq:cutset} with equality.

We will define $\cC$ by its parity-check equations, and we begin with some notation. Let $\{\lambda_{ij}\}_{i\in[n],j\in\{0,1,\dots,s-1\}}$ be $sn$ distinct elements of the field $F$.
Let $g$ be a bijection between the set of pairs $\{(i_1,i_2):i_1,i_2\in[n],i_1<i_2\}$ and the set $\{1,2,\dots,m\}$
defined in \eqref{eq:bn}.
For every $a=0,1,2,\dots,l-1$, we write its expansion in the base $(s^2-1)$ as
 $a=(a_m,a_{m-1},\dots,a_1)$, i.e., $a=\sum_{j=0}^{m-1}a_{j+1}(s^2-1)^j$.
Define the following function 
\begin{equation}\label{eq:deffnew}
\begin{aligned}
f:&\,[n]\times\{0,1,\dots,l-1\}\to \{0,1,\dots,s-1\}\\
&(i,a)\mapsto \Big(\sum_{j=1}^{i-1} b_2(a_{g(j,i)}) + \sum_{j=i+1}^n
b_1(a_{g(i,j)}) \Big) \Mod s,
\end{aligned}
\end{equation}
where $b_1(x)$ and $b_2(x)$ form the digits of the expansion of $x$ in the base $s$; see definition \eqref{eq:defb}.
Note that when $d=k+1$, the function $f$ defined in \eqref{eq:deffnew} is the same as the function defined in \eqref{eq:deff} in Section~\ref{sect:warmup}.
\begin{definition}\label{def:ex}
The code $\cC=\cC_{2,d}$ is defined by the following $rl$ parity check equations.
$$
\sum_{i=1}^n \lambda_{i,f(i,a)}^t c_{i,a} = 0,\;
t=0,1,2,\dots,r-1 , \, a=0,1,2,\dots,l-1.
$$
\end{definition}
For a given $a=0,1,\dots,l-1$ the set of vectors $\{(c_{1,a},c_{2,a},\dots,c_{n,a})\}$ forms an MDS code of length $n$ and dimension $k.$ 
Therefore $\cC$ is indeed an $(n,k,l)$ MDS array code.
Also note that when $d=k+1$, the code $\cC$ is the same as the code defined in Section~\ref{sect:warmup}.

Next we show that $\cC$ has optimal repair bandwidth for repairing any two failed nodes from any $d$ helper nodes.
We need several elements of notation which are similar to the notation used in the previous sections.
For $a=0,1,\dots,l-1$, $j\in[m],$ and $u\in\{0,1,2,\dots,s^2-2\}$, let 
$a(j,u):=(a_m,\dots,a_{j+1},u,a_{j-1},\dots,a_1)$.
 For $a=0,1,\dots,l-1$ and $i\in[n]$, we define 
\begin{align*}
\mu_{i,i_1}^{(a,v_2)}:=\sum_{v_1=0}^{s-1} c_{i,a(g_{12},sv_2+v_1)},\; v_2=0,1,\dots,s-2, \\
\mu_{i,i_2}^{(a,v_1)}:=\sum_{v_2=0}^{s-1} c_{i,a(g_{12},sv_2+v_1)},\; v_1=0,1,\dots,s-2,
\end{align*}
where for brevity we again write $g_{12}$ instead of $g(i_1,i_2).$
The following lemma implies that $\cC$ is an MDS code with the $(2,d)$ optimal repair property.
\begin{lemma}\label{lem:fn} Let the failed nodes be $C_{i_1}$ and $C_{i_2},$ $1\le i_1<i_2\le n$ and let $\cR\subset[n],|\cR|=d$
be a set of $d$ helper nodes.
For any $a\in\{0,1,\dots,l-1\}$ and any $v_2\in\{0,1,\dots,s-2\}$, the values  
$\{c_{i_1,a(g_{12},sv_2+v_1)}, v_1=0,1,\dots,s-1\}$
 and $\mu_{i_2,i_1}^{(a,v_2)}$  are uniquely determined by the set of values $\{\mu_{i,i_1}^{(a,v_2)}:i\in \cR\}$.
Similarly, for any $v_1\in\{0,1,\dots,s-2\}$, the values  
$\{c_{i_2,a(g_{12},sv_2+v_1)}, v_2=0,1,\dots,s-1\}$
 and $\mu_{i_1,i_2}^{(a,v_1)}$  are uniquely determined by the set of values $\{\mu_{i,i_2}^{(a,v_1)}:i\in \cR\}$.
\end{lemma}
\begin{IEEEproof}
The parity-check equations that correspond to the row indices $a(g_{12},0),a(g_{12},1),\linebreak[4]
\dots,a(g_{12},s^2-2)$ are as follows:
\begin{equation}\label{eq:s2}
\sum_{i=1}^n \lambda_{i,f(i,a(g_{12},u))}^t c_{i,a(g_{12},u)} = 0,
\;t=0,1,2,\dots,r-1,\,u=0,1,\dots,s^2-2.
\end{equation}
According to definition of the function $f$ in \eqref{eq:deffnew}, if $i\ne i_1,i_2$ then the value of $f$ does not 
depend on the value of the digit $a_{g_{12}}$. Thus, we have
   $$
f(i,a(g_{12},0)) = f(i,a(g_{12},1)) = \dots = f(i,a(g_{12},s^2-2)),
\; i\in[n]\setminus\{i_1,i_2\}.
   $$
Again according to \eqref{eq:deffnew}, for all $u=0,1,2,\dots,s^2-2$, we have
\begin{equation}\label{eq:dz}
\begin{aligned}
f(i_1,a(g_{12},u)) = \big( f(i_1,a(g_{12},0)) + b_1(u) \big) \mod s, \\
f(i_2,a(g_{12},u)) = \big( f(i_2,a(g_{12},0)) + b_2(u) \big) \mod s.
\end{aligned}
\end{equation}
Therefore, we are justified in using the following notation:
   \begin{align}
\lambda_i&:=\lambda_{i,f(i,a(g(i_1,i_2),0))} = \lambda_{i,f(i,a(g(i_1,i_2),1))}
=\lambda_{i,f(i,a(g(i_1,i_2),2))},\; i\not\in\{i_1,i_2\}  \label{eq:rh1}\\
\lambda_{i_1,v}' &:= \lambda_{i_1,v\oplus f(i_1,a(g_{12},0))},\quad
\lambda_{i_2,v}' := \lambda_{i_2,v\oplus f(i_2,a(g_{12},0))},
  \;v\in\{0,1,\dots,s-1\} \nonumber
   \end{align}
where $\oplus$ is addition modulo $s$.
By \eqref{eq:dz}, for every $u=0,1,2,\dots,s^2-2$, we have
\begin{equation}\label{eq:rh2}
\begin{aligned}
\lambda_{i_1,f(i_1,a(g_{12},u))} = \lambda_{i_1,b_1(u)\oplus f(i_1,a(g_{12},0))}=
\lambda_{i_1,b_1(u)}'; \\
\lambda_{i_2,f(i_2,a(g_{12},u))} = \lambda_{i_2,b_2(u)\oplus f(i_2,a(g_{12},0))} = \lambda_{i_2,b_2(u)}'.
\end{aligned}
\end{equation}
Notice that
$$
\{\lambda_{i,0}',\lambda_{i,1}',\dots,\lambda_{i,s-1}'\} = \{\lambda_{i,0},\lambda_{i,1},\dots,\lambda_{i,s-1}\} \text{~for~} i\in\{i_1,i_2\},
$$
and that
$$
\lambda_i\in\{\lambda_{i,0},\lambda_{i,1},\dots,\lambda_{i,s-1}\}
\text{~for all~} i\in[n]\setminus\{i_1,i_2\}.
$$
Therefore $\lambda_{i_1,0}',\lambda_{i_1,1}',\dots,\lambda_{i_1,s-1}',\lambda_{i_2,0}',
\lambda_{i_2,1}',\dots,\lambda_{i_2,s-1}',
\lambda_i,i\in[n]\setminus\{i_1,i_2\}$ are all distinct.
Using \eqref{eq:rh1} and \eqref{eq:rh2}, we can write \eqref{eq:s2} as
\begin{align*}
(\lambda_{i_1,b_1(u)}')^t c_{i_1,a(g_{12},u)} + (\lambda_{i_2,b_2(u)}')^t c_{i_2,a(g_{12},u)}
+ \sum_{i\in[n]\setminus\{i_1,i_2\}} \lambda_i^t c_{i,a(g_{12},u)} = 0 \\
 t=0,1,2,\dots,r-1,\; u=0,1,\dots,s^2-2.
\end{align*}
These equations have exactly the same form as the equations in \eqref{eq:asj}. 
Therefore the remainder of the proof of this lemma follows the steps in the proof of Lemma~\ref{lem:jj}, and 
there is no need to reproduce them here.
\end{IEEEproof}
This lemma enables us to set up a repair procedure for the nodes $C_{i_1}$ and $C_{i_2}$.
 In the first round of repair, $C_{i_1}$ downloads the set of elements
     \begin{equation}\label{eq:ccd}
\bigcup_{v_2=0}^{s-2}\{\mu_{i,i_1}^{(a,v_2)}:a_{g_{12}}=0\}
     \end{equation}
from each helper node $C_i,i\in \cR.$ In the same way, $C_{i_2}$ downloads the set of elements
$$
\bigcup_{v_1=0}^{s-2}\{\mu_{i,i_2}^{(a,v_1)}:a_{g_{12}}=0\}
$$
 from each helper node $C_i,i\in \cR$.
For future use, let us calculate the number of symbols that $C_{i_1}$ downloads from $C_i,i\in \cR,$ i.e., the cardinality of the set in \eqref{eq:ccd}. Since each digit of $a$ in its $(s^2-1)$-ary expansion can take $s^2-1$ possible values, $|\{\mu_{i,i_1}^{(a,v_2)}:a_{g_{12}}=0\}|=l/(s^2-1)$. The set in \eqref{eq:ccd} is the union of $s-1$ such sets, so its cardinality is $(s-1)l/(s^2-1)=l/(s+1)$.
 
According to Lemma~\ref{lem:fn}, after the first round, $C_{i_1}$ knows the values of
   \begin{equation}\label{eq:21}
\Big(\bigcup_{v_2=0}^{s-2}\bigcup_{v_1=0}^{s-1}\{c_{i_1,a(g_{12},sv_2+v_1)}:a_{g_{12}}=0\}\Big) \bigcup \Big(\bigcup_{v_2=0}^{s-2}\{\mu_{i_2,i_1}^{(a,v_2)} : a_{g_{12}}=0\}\Big),
   \end{equation}
and $C_{i_2}$ knows the values of 
   \begin{equation}\label{eq:22}
\Big(\bigcup_{v_1=0}^{s-2}\bigcup_{v_2=0}^{s-1}\{c_{i_2,a(g_{12},sv_2+v_1)}:a_{g_{12}}=0\}\Big) \bigcup \Big(\bigcup_{v_1=0}^{s-2}\{\mu_{i_1,i_2}^{(a,v_1)} :a_{g_{12}}=0\}\Big).
   \end{equation}
In the second round of the repair process, the nodes $C_{i_1},C_{i_2}$ exchange the second terms in \eqref{eq:21}-\eqref{eq:22}: namely, 
$C_{i_1}$ downloads the elements in the set $\cup_{v_1=0}^{s-2}\{\mu_{i_1,i_2}^{(a,v_1)}: a_{g_{12}}=0\}$ from $C_{i_2}$, and
$C_{i_2}$ downloads the elements in the set $\cup_{v_2=0}^{s-2}\{\mu_{i_2,i_1}^{(a,v_2)}: a_{g_{12}}=0\}$ from $C_{i_1}$.
After the second round, $C_{i_1}$ knows the values of all the elements in the set
$$
\{c_{i_1,a(g_{12},u)}:a_{g_{12}}=0,u\in\{0,1,2,\dots,s^2-2\}\}
=\{c_{i_1,a}:a\in\{0,1,2,\dots,l-1\}\},
$$
and $C_{i_2}$ knows the values of all the elements in the set
$$
\{c_{i_2,a(g_{12},u)}:a_{g_{12}}=0,u\in\{0,1,2,\dots,s^2-2\}\}
=\{c_{i_2,a}:a\in\{0,1,2,\dots,l-1\}\},
$$
i.e., both $C_{i_1}$ and $C_{i_2}$ have recovered all their coordinates. 
Moreover, in the course of the repair process, $C_{i_1}$ downloads 
$l/(s+1)$ symbols of $F$ from each of the nodes $C_i,i\in \cR\cup\{i_2\}$, and $C_{i_2}$
 downloads $l/(s+1)$ symbols of $F$ from each of the nodes $C_i,i\in \cR\cup\{i_1\}$.
Therefore the total repair bandwidth is $2(d+1)l/(s+1)$, meeting the cut-set bound \eqref{eq:cutset} with equality.

\subsection{Optimal repair of two erasures from arbitrary number of helper nodes}\label{sect:hew}
In this section, we point out a technique which has been used extensively but somewhat implicitly in the literature, and we use it 
to construct $(n,k)$ MDS array codes with the universal $(2,d)$-optimal repair property for all $k\le d\le n-2$ simultaneously.
We only aim to convey the main ideas underlying the universal constructions, and we will not discuss all the details in a rigorous way which would require developing new notation, and would lead to tedious and redundant presentation. The initial
idea to use the expansion of the row index is due to \cite{Cadambe11,Tamo13}, and it was used in \cite{Ye16} to construct
explicit universal families of regenerating codes for centralized repair.
 
To illustrate this technique, let us start from the simplest case of repairing single erasure. Returning to the $(n,k,s=d+1-k)$ MDS code defined by the parity-check equations  in \eqref{eq:org}, we observe that the proof of Lemma~\ref{lem:bb} gives a repair scheme of the first node relying on downloading a $\frac1{s}$ proportion of  symbols from each of the $d$ helper nodes (it also gives the $\mu_i$'s which at this point we ignore). Moreover, as already remarked, with straightforward changes to the construction we can obtain a code
with optimal repair of the $i$th node for any given $i=1,\dots,n.$ Denote this code by $\cC_i.$

The next step is to show how two codes of this kind can be combined to construct an $(n,k,l=s^2)$ MDS code that supports optimal
repair of each of the first two nodes from any $d$ helper nodes. For instance, take the codes $\cC_1,\cC_2$ defined over a field $F$
of size at least $n+2s-2,$ and let $\lambda_{1,0},\lambda_{1,1},\dots,\linebreak[3]\lambda_{1,s-1}, \lambda_{2,0},\lambda_{2,1},\dots,\lambda_{2,s-1}, \lambda_3,\lambda_4,\dots,\lambda_n$ be distinct elements of $F$. 
Define an $(n,k,s^2)$ MDS array code $\cC=\cC_1\odot\cC_2$ over $F$ by the following $rs^2$ parity-check equations:
\begin{equation}\label{eq:pol}
\lambda_{1,a_1}^t c_{1,a} + \lambda_{2,a_2}^t c_{2,a} + \sum_{i=3}^n \lambda_i^t c_{i,a} = 0, \quad
a=0,1,\dots,s^2-1, \quad t=0,1,\dots,r-1,
\end{equation}
where $(a_1,a_2)$ is the two-digit $s$-ary expansion of the row index $a\in\{0,1,\dots,s^2-1\}$.
For the repair of the first node, we fix $a_2$ and let $a_1$ take all the values in the set $\{0,1,\dots,s-1\}$. In this way we divide the coordinates of each node into $s$ groups according to the value of $a_2$, and the parity check equations 
that correspond to each group have exactly the same structure as \eqref{eq:org}. Therefore we can optimally repair the first node from any $d$ helper nodes. At the same time, fixing $a_1$ and varying $a_2$, we can optimally repair the second node in the same way.

It is clear that the code $\cC$ defined by \eqref{eq:pol} is obtained by a combination of the codes $\cC_1$ and $\cC_2$
which is similar to the so-called serial concatenation \cite{BDMP98}.
Now it is easily seen that the code $\cC_{1,d}:=\cC_1 \odot \cC_2 \odot \dots \odot \cC_n$ has the $(1,d)$-optimal repair property. In fact, this code family already appeared in the literature; see Construction 2 in \cite{Ye16}.

Now let us consider cooperative repair of two erasures. For $\cF\subseteq[n], |\cF|=2$ and $k\le d\le n-2$, let $\cC_{\cF,d}$ be the $(n,k,l=s^2-1)$ MDS array code that can optimally repair the failed nodes $C_i,i\in\cF$ from any $d$ helper nodes. Note that $\cC_{\{1,2\},d}$ is the code defined by \eqref{eq:asj}, and we previously denoted it as $\cC_{2,d}^{(0)}$. 
As before, the specific choice of $\cF$ is not important, and we can construct a code $\cC_{\cF,d}$ 
with the same structure and parameters as $\cC_{\{1,2\},d}$ for any $2$-subset $\cF\subset [n].$
Now it is clear that the code $\cC_{2,d}$ in Definition~\ref{def:ex} is the concatenation of all $\cC_{\cF,d}$ such that $\cF\subseteq[n], |\cF|=2$, i.e., 
$$
\cC_{2,d} = \bigodot\limits_{\cF\subseteq[n], |\cF|=2} \cC_{\cF,d}.
$$
Following this line of thought, we can easily construct an $(n,k)$ MDS array code $\cC_2^U$ with the {\em universal $(2,d)$-optimal repair property} for all $k\le d\le n-2$ simultaneously. Namely, the concatenated code\footnote{{It is easy to see that the code $\cC_{2,n-2}$ has the $(2,d)$-optimal repair property not only for $d=n-2,$ but also for $d=k.$ Therefore in the concatenation we do not need to include $\cC_{2,k}$.}}
   $$
\cC_2^U := \bigodot\limits_{k+1\le d\le n-2} \cC_{2,d}
   $$
can optimally repair any two failed nodes from any subset of $d$ helper nodes as long as $d\ge k$. The size of the finite field is
determined by the code $\cC_{2,n-2}$ and is at least $(r-1)n$, and the sub-packetization of the code $\cC_2^U$ equals
    $
    \prod_{d=k+1}^{n-2}\big((d-k+1)^2-1\big)^{\binom n2}.
    $

\section{Cooperative $(h,k+1)$ optimal codes for general $h$}\label{sect:h}
\subsection{Repairing the first $h$ nodes from any $d=k+1$ helper nodes} \label{sect:gh}
In this section we present a construction of MDS array codes that can optimally repair the first $h$ nodes from any $d=k+1$ 
helper nodes for any given $h=2,\dots,r-1$. More specifically, given any $k<n,$ any $h\le r-1,$ and a finite field $F$ of
cardinality $|F|\ge n+h$, we present an $(n,k,h+1)$ MDS array code $\cC=\cC_{h,k+1}^{(0)}$ over the field $F$ that has the following property.
When the first $h$ nodes of $\cC$ fail, the repair of each failed node can be accomplished by connecting to {\em any} $k+1$ helper nodes and downloading $k+h$ symbols of $F$ in total from these helper nodes as well as from other failed nodes.
Clearly, the amount of downloaded data meets the cut-set bound \eqref{eq:cutset} with equality.

Let $(\lambda_{ij},i=1,\dots,h, j=0,1), 
\lambda_{h+1},\lambda_{h+2},\dots,\lambda_n$ be $n+h$ distinct elements of the field $F$.
The code $\cC$ is defined by the following parity check equations.
\begin{equation}\label{eq:eov}
\begin{aligned}
\sum_{i=1}^h\lambda_{i,0}^t c_{i,0} + \sum_{i=h+1}^n \lambda_i^t c_{i,0} & = 0,\; t=0,1,\dots,r-1;  \\
\lambda_{a,1}^t c_{a,a} + \sum_{i\in[h]\setminus\{a\}} \lambda_{i,0}^t c_{i,a} + \sum_{i=h+1}^n \lambda_i^t c_{i,a} & = 0,\; t=0,1,\dots,r-1,\, a=1,2,\dots,h.
\end{aligned}
\end{equation}
For every $a=0,1,\dots,h,$ the set of vectors $\{(c_{1,a},c_{2,a},\dots,c_{n,a})\}$ forms an $(n,k)$ MDS code,
therefore $\cC$ is indeed an $(n,k,h+1)$ MDS array code.
When $h=2$, this code is the same as the code defined in Section~\ref{sect:bdblock}.

For $i\in[n]$ and $j\in[h]$, define 
$$
\mu_{ij}:=c_{i,0}+c_{ij}.
$$
Similarly to the previous sections, we have the following lemma:
\begin{lemma}\label{lem:gh} Let $C_1,\dots,C_h$ be the failed nodes.
For any set of helper nodes $\cR\subseteq \{h+1,h+2,\dots,n\},|\cR|=k+1$ and any $j\in[h]$, the values of $c_{j,0},c_{j,j}$ and the sums $\{\mu_{ij},i\in[h]\setminus\{j\}\}$ are uniquely determined by $\{\mu_{ij}:i\in \cR\}$.
\end{lemma}
The proof of this lemma is the same as that of Lemma~\ref{lem:bb}, and we do not repeat it here. 
This lemma implies that the first $h$ nodes of $\cC$ can be repaired with optimal bandwidth.
In the first round, every failed node $C_j,j\in[h]$ downloads $\mu_{ij}$ from each helper node $C_i,i\in \cR$.
According to Lemma~\ref{lem:gh}, after the first round, for every $j\in[h]$, the node $C_j$ knows the values of
$c_{j,0},c_{j,j}$ and $\{\mu_{ij},i\in[h]\setminus\{j\}\}$.
In the second round, every failed node $C_j,j\in[h]$ downloads the sum $\mu_{ji}$ from each of the other failed nodes $C_i,i\in[h]\setminus\{j\}$. After the second round, every failed node $C_j,j\in[h]$ knows the values of $c_{j,0},c_{j,j}$ and the sums $c_{j,0}+c_{j,i},i\in[h]\setminus\{j\}$. Therefore $C_j$ can recover all its coordinates. Moreover, in the whole repair process, every failed node $C_j,j\in[h]$ downloads only one symbol of $F$ from each of the nodes $C_i,i\in \cR\cup [h]\setminus\{j\}$.
Therefore the total repair bandwidth is $h(k+h)$, meeting the cut-set bound \eqref{eq:cutset} with equality.

\subsection{Repairing arbitrary $h$ nodes}\label{sect:lo}

In this section we construct explicit MDS array codes that support $(h,k+1)$-optimal repair of any
$h$-tuple of failed nodes. 
More specifically, given any $k<n,$ any $h\le r-1,$ and a finite field $F$ of cardinality $|F|\ge 2n$, 
we present an $(n,k,l=(h+1)^m)$ MDS array code $\cC=\cC_{h,k+1}$ over the field $F$, where $m:=\binom{n}{h}$. The code $\cC$ has
the property that for {\em any} $h$-subset $\cF$ of $[n],$ the repair of each failed node $C_i,i\in\cF$ can be accomplished by connecting 
to {\em any} $k+1$ helper nodes and downloading $(k+h)l/(h+1)$ symbols of $F$ in total from these helper nodes as well as from other failed nodes.
Clearly, the amount of downloaded data meets the cut-set bound \eqref{eq:cutset} with equality.

As in the previous sections, we will define $\cC$ by its parity-check equations, and we begin with some notation. Let $\{\lambda_{ij}\}_{i\in[n],j\in\{0,1\}}$ be $2n$ distinct elements of the field $F$.
Let $g$ be a bijection between the set of $h$-subsets $\{\cF:\cF\subseteq [n],|\cF|=h\}$ and the numbers $\{1,2,\dots,m\}.$
As in \eqref{eq:bn}, the particular choice of $g$ does not matter; for instance, we can take 
\begin{equation}\label{eq:Dg}
   g(\{i_h,i_{h-1},\dots,i_1\})=\sum_{j=0}^{h-1}\binom{i_{h-j}-1}{h-j}+1
\text{~~for all~} n\ge i_h > i_{h-1} >\dots >i_1\ge 1,
\end{equation}
where we use the convention that $\binom{n_1}{n_2}=0$ if $n_1<n_2$.
For a given $a=0,1,2,\dots,l-1$, let $a_m,a_{m-1},\dots,a_1$ be the digits of its expansion in the base $h+1,$ i.e., $a=\sum_{j=0}^{m-1}a_{j+1}(h+1)^j$. 
For a set $\cF\subseteq[n]$ and an element $i\in\cF$, let 
$z(\cF,i)=|\{j:j\in\cF,j\le i\}|$ be the number of elements in $\cF$ that are no larger than $i$.
Define the following function:
\begin{equation}\label{eq:lo}
\begin{aligned}
    f:\,&[n]\times\{0,1,\dots,l-1\}\to\{0,1\}\\
&(i,a)\mapsto\Big(\sum_{\cF\subseteq [n],|\cF|=h,\;\cF\ni\, i} \mathbbm{1}\{a_{g(\cF)}=z(\cF,i)\}  \Big) \Mod 2,
\end{aligned}
\end{equation}
where $\mathbbm{1}$ is the indicator function.
Finally, given $a=0,1,\dots,l-1,\,i\in[m]$ and $u=0,1,2,\dots,h$, let 
$a(i,u):=(a_m,\dots,a_{i+1},u,a_{i-1},\dots,a_1).$

\begin{definition}
The code $\cC=\cC_{h,k+1}$ is defined by the following $rl$ parity-check equations:
$$
\sum_{i=1}^n \lambda_{i,f(i,a)}^t c_{i,a} = 0
,\; t=0,1,2,\dots,r-1;\, a=0,1,2,\dots,l-1.
$$
\end{definition}
For a given $a=0,1,2,\dots,l-1$ the vectors $(c_{1,a},c_{2,a},\dots,c_{n,a})$ form an $(n,k)$ MDS code.
Therefore $\cC$ is indeed an $(n,k,l)$ MDS array code.

Let us show that $\cC$ has the $(h,k+1)$-optimal repair property.
As before, we define sums of particular entries of the $i$th node. Namely, let $\cF=\{i_1,i_2,\dots,i_h\}$, where $i_1<i_2<\dots<i_h$, be an $h$-subset of $[n].$
Given $a=0,1,\dots,l-1,j\in[h]$ and $i\in[n]$, let
$$
\mu_{i,i_j}^{(a)}:=c_{i,a(g(\cF),0)}+c_{i,a(g(\cF),j)}.
$$
The following lemma implies the optimal bandwidth of $\cC$ for repairing $h$ failed nodes.
\begin{lemma}\label{lem:lo} Let $\cF=\{i_1,i_2,\dots,i_h\}$ be the set of failed nodes.
For any set of helper nodes $\cR\subseteq [n]\setminus\cF,|\cR|=k+1$, any $j\in[h],$ and any $a\in\{0,1,\dots,l-1\}$,
the values of $c_{i_j,a(g(\cF),0)},c_{i_j,a(g(\cF),j)}$ and $\{\mu_{i,i_j}^{(a)}:  i\in\cF\setminus\{i_j\}\}$ are uniquely determined by $\{\mu_{i,i_j}^{(a)}:i\in \cR\}$.
\end{lemma}
The proof of this lemma relies on the same ideas as the proofs of Lemmas~\ref{lem:cv} and \ref{lem:fn}. For completeness we outline it at the end of this section.

Let us explain why Lemma~\ref{lem:lo} implies that $C_i,i\in\cF$ can be repaired with optimal bandwidth.
 In the first round of the repair process, every failed node $C_{i_j},j\in[h]$ downloads 
$\{\mu_{i,i_j}^{(a)}:a_{g(\cF)}=0\}$ from each helper node $C_i,i\in \cR$.
According to Lemma~\ref{lem:lo}, after the first round, $C_{i_j}$ knows the values of
   $$
   \{c_{i_j,a}:a_{g(\cF)}=0\}\cup\{c_{i_j,a(g(\cF),j)}:a_{g(\cF)}=0\}
\cup\{c_{i,a}+c_{i,a(g(\cF),j)}:a_{g(\cF)}=0,i\in\cF\setminus\{i_j\}\}.
   $$
In the second round of the repair process, every failed node $C_{i_j},j\in[h]$ downloads $\{c_{i_j,a}+c_{i_j,a(g(\cF),j')}:a_{g(\cF)}=0\}$ from each of the other failed nodes $C_{i_{j'}},j'\in[h]\setminus\{j\}$. As a result, 
$C_{i_j}$ knows the values of all the elements in the set
   $$
\{c_{i_j,a(g(\cF),u)}:a_{g(\cF)}=0,u=0,1,\dots,h\}
=\{c_{i_j,a}:a\in\{0,1,2,\dots,l-1\}\},
   $$
or, in other words, $C_{i_j}$ can recover all its coordinates. In regards to the repair bandwidth expended during
the two rounds of communication, every failed node $C_{i_j},j\in[h]$ downloads $l/(h+1)$ symbols of $F$ from each of the nodes $C_i,i\in \cR\cup\cF\setminus\{i_j\}$.
Therefore the total repair bandwidth is $h(k+h)l/(h+1)$, meeting the cut-set bound \eqref{eq:cutset} with equality.

\vspace*{0.1in}{\em Proof of Lemma~\ref{lem:lo}:}
The parity-check equations that correspond to the rows labeled by
$a(g(\cF),0),\linebreak[4] a(g(\cF),1),\dots,a(g(\cF),h)$ are as follows:
\begin{equation}\label{eq:eli}
\sum_{i=1}^n \lambda_{i,f(i,a(g(\cF),u))}^t c_{i,a(g(\cF),u)} = 0,\;
t=0,1,2,\dots,r-1,\, u=0,1,2,\dots,h.
\end{equation}
According to definition of the function $f$ in \eqref{eq:lo}, if $i\not\in \cF,$ then the value of $f(i,a)$ does not
depend on the digit of $a$ in position $g(\cF).$ Thus we have
    $$
f(i,a(g(\cF),0)) = f(i,a(g(\cF),1)) = \dots = f(i,a(g(\cF),h)),\; i\in[n]\setminus\cF.
    $$
Likewise we have for any $j\in[h]$
   \begin{align*}
   f(i_j,a(g(\cF),0)) &\neq f(i_j,a(g(\cF),j)),\\
   f(i_j,a(g(\cF),0)) &= f(i_j,a(g(\cF),j')), \;j'\in[h]\backslash\{j\}.
   \end{align*}
Thus we are justified in using the following notation:
  \begin{align}
  \lambda_i&:=\lambda_{i,f(i,a(g(\cF),0))} = \lambda_{i,f(i,a(g(\cF),1))} = \dots
=\lambda_{i,f(i,a(g(\cF),h))},\;i\in[n]\backslash\cF ; \label {eq:ol1}\\
  &\begin{array}{l}
  \lambda_{i_j,0}'  := \lambda_{i_j,f(i_j,a(g(\cF),0))} = \lambda_{i_j,f(i_j,a(g(\cF),j'))}
, j\in[h],\, j'\in[h]\setminus\{j\} ;\\[.1in]
\lambda_{i_j,1}'  := \lambda_{i_j,f(i_j,a(g(\cF),j))},\;j\in[h].
  \end{array}\label{eq:ol2}
  \end{align}
%
Notice that
    \begin{gather*}
\lambda_{i_j,0}'\neq \lambda_{i_j,1}' \text{~and~}
\{\lambda_{i_j,0}', \lambda_{i_j,1}'\}=\{\lambda_{i_j,0}, \lambda_{i_j,1}\}
\text{~for all~} j\in[h], \\
\lambda_i\in\{\lambda_{i,0},\lambda_{i,1}\},\; i\in[n]\setminus\cF.
    \end{gather*}
Therefore the elements $\lambda_{i_1,0}',\lambda_{i_2,0}',\dots,\lambda_{i_h,0}',
\lambda_{i_1,1}',\lambda_{i_2,1}',\dots,\lambda_{i_h,1}',
\lambda_i,i\in[n]\setminus\cF$ are all distinct.
Now we can write \eqref{eq:eli} as
\begin{align*}
\sum_{j=1}^h (\lambda_{i_j,0}')^t c_{i_j,a(g(\cF),0)} + \sum_{i\in[n]\setminus\cF} \lambda_i^t c_{i,a(g(\cF),0)}  = 0 ,\; t=0,1,\dots,r-1;  \\
(\lambda_{i_u,1}')^t c_{i_u,a(g(\cF),u)} + \sum_{j\in[h]\setminus\{u\}} (\lambda_{i_j,0}')^t c_{i_j,a(g(\cF),u)} + \sum_{i\in[n]\setminus\cF} \lambda_i^t c_{i,a(g(\cF),u)}  = 0 \\ t=0,1,\dots,r-1;\; u=1,2,\dots,h.
\end{align*}
These equations have exactly the same form as the equations in \eqref{eq:eov}. 
Therefore the remainder of the proof of  Lemma~\ref{lem:lo} follows the steps in the proof of Lemma~\ref{lem:gh} (or Lemma~\ref{lem:bb}), 
and we do not repeat them here.

\section{Cooperative $(h,d)$-optimal codes for general $h$ and general $d$}\label{sect:fg}
\subsection{Repairing the first $h$ nodes from any $d$ helper nodes} \label{sect:hd0}
In this section we present a construction of MDS array codes that can optimally repair the first $h$ nodes from any $d\ge k+1$ helper nodes for any given $2\le h\le n-d\le r-1.$ (We do not consider the case of $d=k$ because codes for it were constructed earlier in \cite{Shum13}.)
Let $s:=d+1-k$.
 Given a finite field $F$ of
cardinality $|F|\ge n+h(s-1)$, we present an $(n,k,l=(h+s-1)(s-1)^{h-1})$ MDS array code $\cC=\cC_{h,d}^{(0)}$ over the field $F$ that has the following property:
When the first $h$ nodes of $\cC$ fail, the repair of each failed node can be accomplished by connecting to {\em any} $d$ helper nodes and downloading 
$$
(d+h-1)\frac{l}{d+h-k} = 
(d+h-1)(s-1)^{h-1}
$$
 symbols of $F$ in total from these helper nodes as well as from the other failed nodes.
Clearly, the amount of downloaded data meets the cut-set bound \eqref{eq:cutset} with equality.

Let $(\lambda_{ij},i=1,\dots,h, j=0,1,\dots,s-1), 
\lambda_{h+1},\lambda_{h+2},\dots,\lambda_n$ be $hs+n-h$ distinct elements of the field $F$.
Define 
\begin{equation}\label{eq:dA}
A:=\{\underline{a}=(a_1,a_2,\dots,a_h) :\underline{a}\in \{0,1,\dots, s-1\}^h, \sum_{i=1}^h \mathbbm{1}\{a_i=s-1\}\le 1\},
\end{equation}
 i.e., $A$ is the subset of $\{0,1,\dots, s-1\}^h$ consisting of all the $\underline{a}$ such that at most one of its coordinates is $s-1$.
It is easy to verify that 
\begin{equation}\label{eq:cardA}
|A|=(h+s-1)(s-1)^{h-1} = l.
\end{equation}

Let $C=(C_1,C_2,\dots,C_n)\in \cC$ be a codeword of the code $\cC$.
In this section, we use a multi-index (vector) notation $\underline{a}=(a_1,a_2,\dots,a_h)$ to label the  entries of each node $C_i$, so 
the node has the form $C_i=(c_{i,\underline{a}},\underline{a}\in A).$ In previous sections we opted for numbering the
entries of $C_i$ with integers even though on several occasions (e.g., in Sections \ref{sect:warmup}, \ref{sect:rb})
we have essentially relied on the multi-index notation. We could follow this pattern in this section as well, however the integer numbering
would not be consecutive, and we find the vector notation much more convenient for the presentation.
We note that, according to \eqref{eq:cardA}, the dimension of $C_i$ over $F$ is indeed $l$.

\begin{definition}
The code $\cC$ is defined by the following parity check equations.
\begin{equation}\label{eq:pcq}
\sum_{i=1}^h \lambda_{i,a_i}^t c_{i,\underline{a}} + \sum_{i=h+1}^n \lambda_i^t c_{i,\underline{a}}=0, \quad t=0,1,\dots,r-1, \quad \underline{a}\in A.
\end{equation}
\end{definition}
Since for each $\underline{a}\in A$, the set of vectors
$\{(c_{1,\underline{a}}, c_{2,\underline{a}},\dots, c_{n,\underline{a}})\}$ forms an $(n,k)$ MDS code, $\cC$ is indeed an $(n,k,l)$ MDS array code.

\subsubsection{Intuition behind the repair scheme}
We begin with an informal discussion of the code construction and the accompanying repair scheme.
According to the cut-set bound \eqref{eq:cutset}, if we assume that the amount of communication between any two nodes is the same ({\em uniform download}), which is the case for our repair scheme, then this amount is equal to $\frac{l}{h+d-k} = (s-1)^{h-1}$ symbols of $F$. More precisely, in the first round of repair process, each failed node should download $(s-1)^{h-1}$ symbols of $F$ from each helper node, and in the second round, each failed node should download $(s-1)^{h-1}$ symbols of $F$ from each of the other failed nodes.

For $i\in[h]$ and $u\in\{0,1,\dots,s-1\}$,
define $\underline{a}(i,u):=(a_1,a_2,\dots,a_{i-1},u,a_{i+1},a_{i+2},\dots,a_h)$.
For $i\in[h]$, define the set of indices
$$
B_i:=\{\underline{a}=(a_1,a_2,\dots,a_h): a_i\in[0,s-1], a_j\in[0,s-2] \text{ for all }j\ne i\},
$$
where $[0,t]:=\{0,1,\dots,t\}$ for an integer $t$.
Define $A_0:=\{0,1,\dots,s-2\}^h$. It is easy to see that 
  $$
\bigcup_{i=1}^h B_i=A,\quad
\bigcap_{i=1}^h B_i=A_0.
  $$
In the first round of repair, each failed node $C_i,i\in[h]$ connects to $d$ helper nodes $C_j,j\in\cR$ and downloads $(s-1)^{h-1}$ symbols from each of them, so altogether it acquires $d (s-1)^{h-1}$ symbols of $F$.
This enables $C_i$ to recover a certain portion of its entries, which we can quantify relying on the cut-set bound.
For this, we observe that this bound gives a lower estimate on the repair bandwidth for a given size of each node $l$. At the same time,
given the repair bandwidth, it gives an upper estimate on the node size, including in particular a 
bound on the maximum number of entires of the node that can be recovered from a certain amount of the downloaded data. 
Using this observation, let us take $|\cF|=1$ and $|\cR|=d$ in \eqref{eq:csce} (or in \eqref{eq:cutset}), and
replace the left-hand side with $d (s-1)^{h-1}$. Solving for $l$, we see that each failed node can recover at most $s(s-1)^{h-1}$ coordinates. 
At the same time, the cardinality of the set $B_i$ is exactly $s(s-1)^{h-1}$, and this is the subset of the  entries of $C_i$ that will be
repaired after the first round of communication. Namely, according to Lemma~\ref{lem:tch}, the set of values
 $\{c_{i,\underline{a}}:\underline{a}\in B_i\}$ can be found relying on the values
$$
\Big\{\Big(\sum_{u=0}^{s-1} c_{j,\underline{a}(i,u)}: \underline{a}\in B_i, a_i = 0 \Big), j\in \cR\Big\}
$$
(see Lemma \ref{lem:sfg} below), and therefore, the node $C_i$ downloads the set $\{\sum_{u=0}^{s-1} c_{j,\underline{a}(i,u)}: \underline{a}\in B_i, a_i = 0\}$
from each of the helper nodes $C_j,j\in\cR$. Since for every $\underline{a}\in B_i$ the coordinate $a_i$ can take $s$ possible values, the number of symbols downloaded from each of them is exactly $\frac{|B_i|}{s}=(s-1)^{h-1}$.

To move forward, we note that Lemma~\ref{lem:tch} gives us more: namely, apart from the values $\{c_{i,\underline{a}}:\underline{a}\in B_i\},$ each $C_i, i\in [h]$ can also compute 
$(s-1)^{h-1}$ {\em sums of coordinates of the other failed nodes}. Namely, after the first round, $C_i$ can find the values
    \begin{equation}\label{eq:kjk}
\Big\{\sum_{u=0}^{s-1} c_{j,\underline{a}(i,u)}: \underline{a}\in B_i, a_i = 0\Big\} \quad \text{for all }j\in[h]\setminus\{i\}.
\end{equation}
This is the information that will be exchanged between the failed nodes $C_i, i\in [h]$ in the second round.
 
To describe the second part of the repair scheme, we note that the number of coordinates still not available at the node $C_i$ equals 
    $$
|A\setminus B_i|=
     l-s(s-1)^{h-1}=(h-1)(s-1)^{h-1}.
    $$
As noted above (again assuming uniform download), in the second round each failed node should download $(s-1)^{h-1}$ symbols of $F$ from each of the other $(h-1)$ failed nodes. Therefore, in the second round, each failed node should acquire $(h-1)(s-1)^{h-1}$ symbols of $F,$ which matches the number of the still missing symbols of the node.
To decide what to download we turn to \eqref{eq:kjk}, noting that 
each failed node $C_i$ knows the sums in \eqref{eq:kjk} for all the other failed nodes $C_j,j\in[h]\setminus\{i\}.$
For a fixed $j$, there are $(s-1)^{h-1}$ symbols in the set \eqref{eq:kjk}, 
so a natural thing to do in the second round is to let $C_i$ transmit the sums in \eqref{eq:kjk} to each
of the remaining failed nodes $C_j,j\in[h]\setminus\{i\}$.

Since every failed node $C_j$ knows $\{c_{j,\underline{a}}:\underline{a}\in B_j\}$ after the first round and $A_0\subset B_j$ for all $j\in[h]$, every failed node $C_j$ knows $\{c_{j,\underline{a}}:\underline{a}\in A_0\}$. We observe that each sum in \eqref{eq:kjk} has $s$ terms and that the indices of $s-1$ of them belong to the set $A_0$, so $C_j$ can calculate the 
single remaining term from each of these sums. Upon completing this calculation, the node $C_j$ 
knows the values of all the summands of all the sums in the set \eqref{eq:kjk}, i.e., $C_j$ knows all the coordinates in the set
$\{c_{j,\underline{a}}:\underline{a}\in B_i\}.$ 
Since $C_j$ downloads these sums from all the other failed nodes $C_i,i\in[h]\setminus\{j\}$, the downloaded symbols in the second round enable $C_j$ to calculate the coordinates 
$$
\bigcup_{i\in[h]\setminus\{j\}} \{c_{j,\underline{a}}:\underline{a}\in B_i\big\}.
$$
Recall that after the first round, $C_j$ already knows the values of coordinates $\{c_{j,\underline{a}}:\underline{a}\in B_j\}$. Thus after the whole repair process, $C_j$ can find the entries
$$
\Big\{c_{j,\underline{a}}:\underline{a}\in \bigcup_{i=1}^h B_i\big\}
=\{c_{j,\underline{a}}:\underline{a}\in A\}.
$$
This concludes the repair procedure because $C_j$ has found all the missing $l$ entries.

\vspace*{.1in}\subsubsection{Formal description and validity proof of the repair scheme}
The discussion in the previous subsection contains most of what is needed to justify the repair scheme.
The omitted step is a connection with Lemma~\ref{lem:tch}
which we include next.

\begin{lemma}\label{lem:sfg}
Let $C_i,i\in[h]$ be one of the failed nodes, and let $\cR\subseteq [n]\setminus[h]$ be the indices of helper nodes, where
$|\cR|=d$.
For any $\underline{a}\in B_i$, the elements $c_{i,\underline{a}(i,0)},c_{i,\underline{a}(i,1)},\dots,c_{i,\underline{a}(i,s-1)}$ and the values of $\{\sum_{u=0}^{s-1} c_{j,\underline{a}(i,u)}:j\in[h]\setminus\{i\}\}$
can be calculated from the values in the set $\{\sum_{u=0}^{s-1} c_{j,\underline{a}(i,u)}:j\in \cR\}$.
\end{lemma}
\begin{IEEEproof}
We again use Lemma~\ref{lem:tch}. Let us write out the parity-check equations \eqref{eq:pcq} that 
correspond to the indices $\underline{a}(i,0), \underline{a}(i,1),\dots, \underline{a}(i,s-1)$:
\begin{align}\label{eq:uson}
\lambda_{i,u}^t c_{i,\underline{a}(i,u)} +
\sum_{j\in[h]\setminus\{i\}} \lambda_{j,a_j}^t c_{j,\underline{a}(i,u)} &+ \sum_{j=h+1}^n \lambda_j^t c_{j,\underline{a}(i,u)}=0,
\nonumber\\  &t=0,1,\dots,r-1, \quad u=0,1,\dots,s-1
\end{align}
We can see that this set of equations has the same form as \eqref{eq:org}: In \eqref{eq:uson} only the coefficients of $c_{i,\underline{a}(i,u)}$ vary with $u$ while the coefficients of $c_{j,\underline{a}(i,u)}$ are independent of $u$ for all $j\in[n]\setminus\{i\}$;
in \eqref{eq:org} only the coefficients of $c_{1,u}$ vary with $u$ while the coefficients of $c_{j,u}$ are independent of $u$ for all $j\in[n]\setminus\{1\}$.
Therefore Lemma~\ref{lem:tch} applies directly, and the proof is complete.
\end{IEEEproof}

In the first round, each failed node $C_i,i\in[h]$ downloads 
\begin{equation}\label{eq:wh}
\Big\{\sum_{u=0}^{s-1} c_{j,\underline{a}(i,u)}: \underline{a}\in B_i, a_i = 0\Big\}
\end{equation}
from each helper node $C_j,j\in\cR$. 
As already explained, the cardinality of the set in \eqref{eq:wh} is $(s-1)^{h-1}$.

According to Lemma~\ref{lem:sfg}, after the first round, each failed node $C_i,i\in[h]$ knows the following field elements:
\begin{align*}
\{c_{i,\underline{a}}:\underline{a}\in B_i\}
\bigcup \Big(\bigcup_{j\in[h]\setminus\{i\}} \Big\{ \sum_{u=0}^{s-1} c_{j,\underline{a}(i,u)}: 
\underline{a}\in B_i, a_i = 0 \Big\} \Big).
\end{align*}

In the second round, each failed node $C_j,j\in[h]$ downloads
$$
\Big\{\sum_{u=0}^{s-1} c_{j,\underline{a}(i,u)}: \underline{a}\in B_i, a_i = 0 \Big\}
$$
from each of the other failed nodes $C_i,i\in[h]\setminus\{j\}$.
According to the arguments above, after the second round each failed node can recover all its coordinates, and the repair bandwidth achieves the cut-set bound \eqref{eq:cutset} with equality.

\subsubsection{Connections with $\cC_{2,d}^{(0)}$ and $\cC_{h,k+1}^{(0)}$}\label{sect:connections}
Let us look back at the codes $\cC_{2,d}^{(0)}$ and $\cC_{h,k+1}^{(0)}$ which are special cases of the above construction (although this may be not immediate to see, which justifies their independent description earlier in the paper).
Namely, the code $\cC_{h,d}^{(0)}$ with $h=2$ becomes the same as $\cC_{2,d}^{(0)},$ 
albeit with a different way of indexing the entries of each node $C_i$, and similarly, letting $d=k+1$ in $\cC_{h,d}^{(0)}$, 
we obtain the code $\cC_{h,k+1}^{(0)}$ with a different way of indexing. 

First, using Table~\ref{table:parameters}, it is immediate to see that the sub-packetization values match.
Now let us verify the easier of the two specializations, checking the case of $h=2.$
Indeed, in this case the set $A$ defined in \eqref{eq:dA} becomes 
$$
A=\{\underline{a}=(a_1,a_2): a_1,a_2\in\{0,1,\dots,s-1\}, (a_1,a_2) \neq (s-1,s-1)\}.
$$
A natural way to transform the multi-index $\underline{a}=(a_1,a_2)$ into an integer index is to use the mapping
$a=a_1+sa_2$. It is clear that the image of $A$ under this mapping is $\{0,1,2,\dots,s^2-2\}$, which is exactly the same as the set of integer indices in Section~\ref{sect:fd}.
One can further check that when $h=2$, the parity check equations of $\cC_{h,d}^{(0)}$ given in \eqref{eq:pcq}
are the same as the parity check equations \eqref{eq:asj} of $\cC_{2,d}^{(0)}$. 

Let us now explain that using $d=k+1$ in the description of the code $\cC_{h,d}^{(0)},$ we obtain $\cC_{h,k+1}^{(0)}.$ 
When $d=k+1$, the set $A$ defined in \eqref{eq:dA} becomes
$$
A=\{\underline{0},e_1,e_2,\dots,e_h\},
$$
where $\underline{0}$ is an all-zero vector of length $h$, and for $i\in[h]$, $e_i$ is the $h$-dimensional vector whose only nonzero coordinate is located at the $i$th position, and this coordinate is $1$.
We map $\underline{0}$ to $0$ and $e_i$ to $i$ for all $i\in[h]$. It is easy to check that under this mapping
the parity-check equations \eqref{eq:pcq} of the code $\cC_{h,d}^{(0)}$ are the same as the parity-check equations
\eqref{eq:eov} of $\cC_{h,k+1}^{(0)}$. 

 \subsection{Repairing any $h$ nodes from any $d$ helper nodes}\label{sect:hd1}
Finally, in this section we present the codes $\cC=\cC_{h,d}$ that address the most general case of the repair problem. 
As above, we let $s:=d+1-k$ and suppose that $F, |F|\ge sn$ is a finite field.
We present an $(n,k,l=((h+s-1)(s-1)^{h-1})^m)$ MDS array code $\cC=\cC_{h,d}$ over $F$, where $m:=\binom{n}{h}$.
The code $\cC$ has the property that for {\em any} $h$-subset $\cF$ of $[n],$ the repair of each failed node $C_i,i\in\cF$ can be accomplished by 
connecting to {\em any} $d$ helper nodes and downloading $(d+h-1)l/(h+s-1)$ symbols of $F$ in total from these helper nodes as well as from the other failed nodes.
Clearly, the amount of downloaded data meets the cut-set bound \eqref{eq:cutset} with equality, and so the code $\cC$ supports optimal repair.

Let $\{\lambda_{ij},i=1,\dots,n, j=0,1,\dots,s-1\}$ be $sn$ distinct elements of the field $F$. We will rely on the 
definition of the set $A$ in \eqref{eq:dA}. To remind ourselves, this is the set of $h$-tuples of integers between $0$ and $s-1$ that contain at most one entry equal to $s-1.$
We use the shorthand notation $[0,i]:=\{0,1,\dots,i\}$ for an integer $i$, and
  define a set of integer vectors $A^{[m]}\subset [0,s-1]^{hm}$ such that
each of the $m$ subvectors is contained in $A$. More specifically, in this section we use $\underline{a}$ to denote an integer vector of length $hm$:
  \begin{equation}\label{eq:a}
  \underline{a}= (\underline{a}^{(1)}, \underline{a}^{(2)},\dots,\underline{a}^{(m)}),
  \end{equation}
where $\underline{a}^{(i)}= (a^{(i)}_1,\dots,a^{(i)}_h) \in[0,s-1]^h.$ Define the set 
    $$
A^{[m]} := \{\underline{a}\in [0,s-1]^{hm}: \underline{a}^{(i)}\in A, i=1,\dots,m\}.
    $$
According to \eqref{eq:cardA}, each $\underline{a}^{(i)}$ can take $(h+s-1)(s-1)^{h-1}$ possible values, so 
\begin{equation}\label{eq:aml}
\big| A^{[m]} \big| = \big((h+s-1)(s-1)^{h-1} \big)^m =l.
\end{equation}
Let $g$ be the bijection between the set of $h$-subsets $\{\cF:\cF\subseteq [n],|\cF|=h\}$ and the numbers $\{1,2,\dots,m\}$ 
defined in \eqref{eq:Dg}.
For a set $\cF\subseteq[n]$ and an element $i\in\cF$, let 
$z(\cF,i)=|\{j:j\in\cF,j\le i\}|$ be the number of elements in $\cF$ that are not greater than $i$.
Define the following function:
\begin{equation}\label{eq:dke}
\begin{aligned}
    f:\,&[n]\times A^{[m]} \to\{0,1,\dots,s-1\}\\
&(i,\underline{a})\mapsto\biggl(\sum_{\cF\subseteq [n],|\cF|=h,\;\cF\ni\, i} 
\underline{a}^{(g(\cF))}_{z(\cF,i)}  \biggl) \Mod s,
\end{aligned}
\end{equation}
Let $C=(C_1,C_2,\dots,C_n)\in \cC$ be a codeword of the code $\cC$. We index the entries of the code $C_i$ using
the multi-index $\underline{a}$ defined above in \eqref{eq:a},
writing $C_i=(c_{i,\underline{a}},\underline{a}\in A^{[m]})$. According to \eqref{eq:aml}, the dimension of $C_i$ over $F$ is indeed $l$.
The last element of notation is as follows: for every $\underline{a}\in A^{[m]}$, $i\in[m]$ and $\underline{b}\in A$, let 
$$
\underline{a}(i,\underline{b}):=(\underline{a}^{(1)},\underline{a}^{(2)},\dots,\underline{a}^{(i-1)},\underline{b},
\underline{a}^{(i+1)},\dots\underline{a}^{(m)}).
$$

\begin{definition}
The code $\cC=\cC_{h,d}$ is defined by the following $rl$ parity-check equations:
\begin{equation}\label{eq:hfg}
\sum_{i=1}^n \lambda_{i,f(i,\underline{a})}^t c_{i,\underline{a}} = 0
,\; t=0,1,2,\dots,r-1;\, \underline{a}\in A^{[m]}.
\end{equation}
\end{definition}
For every $\underline{a}\in A^{[m]}$, the vectors $(c_{1,\underline{a}},c_{2,\underline{a}},\dots,c_{n,\underline{a}})$ form an $(n,k)$ MDS code.
Therefore $\cC$ is indeed an $(n,k,l)$ MDS array code.

Let us show that $\cC$ has the $(h,d)$-optimal repair property.
Let $\cF=\{i_1,i_2,\dots,i_h\}$, where $1\le i_1<i_2<\dots<i_h \le n$, be the set of indices of $h$ failed nodes.
For every codeword $C=(C_1,C_2,\dots,C_n)\in \cC$ and every $\underline{a}\in A^{[m]}$, 
we form a vector $C^{(\underline{a})}$ by taking a subset of coordinates from each node $C_i,i\in[n]$:
$$
C^{(\underline{a})}:=(C_1^{(\underline{a})},C_2^{(\underline{a})},\dots,C_n^{(\underline{a})} ),
$$
where 
\begin{equation}\label{eq:iod}
C_i^{(\underline{a})}:=(c_{i,\underline{a}(g(\cF),\underline{b})}: \underline{b}\in A), \quad i=1,\dots,n.
\end{equation}
By definition the set $C_i^{(\underline{a})}$ contains $(h+s-1)(s-1)^{h-1}$ coordinates of $C_i$.
Since the indices of these coordinates are obtained by replacing the subvector $\underline{a}^{(g(\cF))}$ with all the vectors of the set $A$, the vectors $C^{(\underline{a})}$ and $C_i^{(\underline{a})}$ do not depend on the original value of $\underline{a}^{(g(\cF))},$ i.e.,
\begin{equation}\label{eq:fje}
C^{(\underline{a})}=C^{(\underline{a}(g(\cF),\underline{b}))} \text{~and~}
C_i^{(\underline{a})}=C_i^{(\underline{a}(g(\cF),\underline{b}))}
\text{~for all~} C\in\cC, i\in[n] \text{~and~} \underline{b}\in A.
\end{equation}

Moreover, consider the following $((h+s-1)(s-1)^{h-1})^{m-1}$ sets of coordinates of $C_i$:
\begin{equation}\label{eq:ho}
\{C_i^{(\underline{a})}:\underline{a}\in A^{[m]}, \underline{a}^{(g(\cF))}=\underline{0}\},
\end{equation}
where we view each vector $C_i^{(\underline{a})}$ defined in \eqref{eq:iod} as a set. Since we are limiting the subvector
$\underline{a}^{(g(\cF))}$ to $0$ while originally it can take $|A|= (h+s-1)(s-1)^{h-1}$ values, the vector
$\underline{a}$ in \eqref{eq:ho} takes
  $$
  \frac{l}{(h+s-1)(s-1)^{h-1}}=((h+s-1)(s-1)^{h-1})^{m-1}
  $$
   possible values. Therefore \eqref{eq:ho} contains $((h+s-1)(s-1)^{h-1})^{m-1}$ distinct sets of coordinates of $C_i$.
This amounts to saying that the sets in \eqref{eq:ho} form a partition of 
 the coordinates of $C_i$.

For every $\underline{a}\in A^{[m]}$, we define an $(n,k,(h+s-1)(s-1)^{h-1})$ MDS array code $\cC^{(\underline{a})}$ as follows:
$$
\cC^{(\underline{a})}:= \{(C_1^{(\underline{a})},C_2^{(\underline{a})},\dots,C_n^{(\underline{a})}): 
C\in\cC\},
$$
where the MDS property and the dimension of $\cC^{(\underline{a})}$ follow directly from the definition of the code $\cC$; see \eqref{eq:hfg}, \eqref{eq:iod}.
To better understand the connection between the code $\cC$ and its subcodes $\cC^{(\underline{a})},\underline{a}\in A^{[m]}$, we 
can view each codeword of $\cC$ as a two-dimensional array of size $l\times n$.
We use multi-index $\underline{a}\in A^{[m]}$ to index each row and $i\in[n]$ to index each column of the codeword.
Each subcode $\cC^{(\underline{a})},\underline{a}\in A^{[m]}$ contains $(h+s-1)(s-1)^{h-1}$ rows of the codewords in $\cC$, and the indices of these $(h+s-1)(s-1)^{h-1}$ rows are in the set 
$\{\underline{a}(g(\cF),\underline{b}): \underline{b}\in A\}$.
From \eqref{eq:fje} it is clear that
$$
\cC^{(\underline{a})}=\cC^{(\underline{a}(g(\cF),\underline{b}))}
\text{~for all~} \underline{b}\in A.
$$
Thus,  the code $\cC$ can be partitioned into $((h+s-1)(s-1)^{h-1})^{m-1}$ subcodes
$$
\{\cC^{(\underline{a})}:\underline{a}\in A^{[m]}, \underline{a}^{(g(\cF))}=\underline{0}\},
$$
and each subcode contains $(h+s-1)(s-1)^{h-1}$ rows of the code $\cC$.
We will show that each of these subcodes has the same structure as the code $\cC_{h,d}^{(0)}$ defined in Section~\ref{sect:hd0}, and can therefore
be optimally repaired.

\begin{lemma}\label{lem:vhc}
For every $\underline{a}\in A^{[m]}$, the $(n,k,(h+s-1)(s-1)^{h-1})$ MDS array code $\cC^{(\underline{a})}$ can optimally repair the failed nodes $C_i^{(\underline{a})},i\in\cF$ from any $d$ helper nodes, i.e., the bandwidth of repairing $C_i^{(\underline{a})},i\in\cF$ from any $d$ helper nodes achieves \eqref{eq:cutset} with equality.
\end{lemma}
\begin{IEEEproof}
%
Our goal is to show that the code $\cC^{(\underline{a})}$ has the same structure as the code $\cC_{h,d}^{(0)}$. Then we can apply the optimal repair scheme for the first $h$ nodes of $\cC_{h,d}^{(0)}$ to the repair of the failed nodes of $\cC^{(\underline{a})}$ whose indices are in $\cF$.

By definition \eqref{eq:dke}, the function $f$ has the following property:
For any $\underline{a}\in A^{[m]}$ and any $\underline{b}=(b_1,b_2,\dots,b_h)\in A$,
\begin{equation}\label{eq:gej}
\begin{aligned}
f(i,\underline{a}(g(\cF),\underline{b})) &= f(i,\underline{a}) \text{~for all~} i\in[n]\setminus \cF, \\
f(i_u,\underline{a}(g(\cF),\underline{b})) &= f(i_u,\underline{a}(g(\cF),\underline{0})) \oplus b_u
\text{~for all~} u\in[h],
\end{aligned}
\end{equation}
where $\underline{0}$ is the all-zero vector of length $h$, and $\oplus$ is addition modulo $s$.
From now on we fix an $\underline{a}\in A^{[m]}$ and prove the claim for this fixed $\underline{a}$.
According to \eqref{eq:gej}, we are justified in using the following notation:
\begin{equation}\label{eq:lkj1}
\lambda_i := \lambda_{i,f(i,\underline{a})} = \lambda_{i,f(i,\underline{a}(g(\cF),\underline{b}))}
\text{~for all~} i\in[n]\setminus \cF \text{~and all~} \underline{b}\in A.
\end{equation}
We further define
$$
\lambda_{i_u,j}' := \lambda_{i_u,f(i_u,\underline{a}(g(\cF),\underline{0})) \oplus j}
\text{~for all~} u\in[h] \text{~and all~} j\in\{0,1,\dots,s-1\}.
$$
Again by \eqref{eq:gej}, we have
\begin{equation}\label{eq:lkj2}
\lambda_{i_u,b_u}' = \lambda_{i_u, f(i_u,\underline{a}(g(\cF),\underline{0})) \oplus b_u}
= \lambda_{i_u,f(i_u,\underline{a}(g(\cF),\underline{b}))}
\text{~for all~} u\in[h] \text{~and all~} \underline{b}\in A.
\end{equation}
By \eqref{eq:iod}, $C_i^{(\underline{a})}$ consists of the coordinates $(c_{i,\underline{a}(g(\cF),\underline{b})}: \underline{b}\in A)$. Using \eqref{eq:hfg}, \eqref{eq:lkj1} and \eqref{eq:lkj2}, we can write out the parity check equations of  $\cC^{(\underline{a})}$ as follows:
\begin{equation}\label{eq:poi}
\sum_{u=1}^h (\lambda_{i_u,b_u}')^t c_{i_u,\underline{a}(g(\cF),\underline{b})} +
\sum_{i\in[n]\setminus\cF} \lambda_i^t c_{i,\underline{a}(g(\cF),\underline{b})} = 0,
\quad  t=0,1,\dots,r-1, \quad \underline{b}\in A.
\end{equation}
We can check that \eqref{eq:poi} has the same form as \eqref{eq:pcq}. Indeed, $\underline{b}$ in \eqref{eq:poi} plays the role of $\underline{a}$ in \eqref{eq:pcq}; the first sum in both equations consists of coordinates of the $h$ failed nodes, 
and the second sum in both equations consists of coordinates of the other available nodes; in both equations, only the coefficients of the coordinates of the failed nodes vary with the indices, and they vary in exactly the same way.
Therefore the repair scheme of code $\cC_{h,d}^{(0)}$ can be directly applied to the repair of $C_i^{(\underline{a})},i\in\cF$ from any $d$ helper nodes, and the repair bandwidth of this scheme achieves the bound \eqref{eq:cutset}. This completes the proof of Lemma~\ref{lem:vhc}.
\end{IEEEproof}
Since every subcode can optimally repair the failed nodes whose indices are in the set $\cF,$
the same is true for the code $\cC$: namely it is capable of repairing $C_i,i\in\cF$ from any $d$ helper nodes with optimal repair bandwidth.

\vspace*{.1in}
{\em Remark:} Expanding the discussion in Section~\ref{sect:connections}, we can see
that both the codes $\cC_{2,d}$ and $\cC_{h,k+1}$ are special cases of the code $\cC_{h,d}:$
taking $h=2$ in the definition of $\cC_{h,d}$, we obtain the code $\cC_{2,d}$ with a different indexing of the node's coordinates, and in the same way, taking $d=k+1$ in $\cC_{h,d}$, we obtain the code $\cC_{h,k+1},$ with a different way of indexing.

\subsection{A family of universal codes}\label{sect:universal}
Using the construction in the previous subsection as a building block and exploiting the concatenation operation defined in Section~\ref{sect:hew}, we can easily construct an $(n,k)$ MDS array code $\cC^U$ with universal $(h,d)$-optimal repair property for all $1\le h\le n-d\le n-k$ simultaneously. In other words, the codes that we construct can optimally repair any number of erasures from any number of helper nodes.

Indeed, let
    $$
    \cC^U := \bigodot_{1\le h\le n-d\le n-k} \cC_{h,d}.
    $$
The code $\cC^U$ is simply a concatenation of all $\cC_{h,d}$ for $1\le h\le n-d\le n-k$, where the codes $\cC_{h,d}$ for $h\ge 2$ 
are defined in the previous subsection, and the code $\cC_{1,d}$ is given in Sec.~\ref{sect:hew} \cite{Ye16}. It can be constructed
over a field $F$ with size $|F|\ge rn,$ and it supports optimal repair of any single node, and optimal cooperative repair of any $h\ge 2$ nodes.

\bibliographystyle{IEEEtran}
\bibliography{repair}

\begin{thebibliography}{10}
\providecommand{\url}[1]{#1}
\csname url@samestyle\endcsname
\providecommand{\newblock}{\relax}
\providecommand{\bibinfo}[2]{#2}
\providecommand{\BIBentrySTDinterwordspacing}{\spaceskip=0pt\relax}
\providecommand{\BIBentryALTinterwordstretchfactor}{4}
\providecommand{\BIBentryALTinterwordspacing}{\spaceskip=\fontdimen2\font plus
\BIBentryALTinterwordstretchfactor\fontdimen3\font minus
  \fontdimen4\font\relax}
\providecommand{\BIBforeignlanguage}[2]{{%
\expandafter\ifx\csname l@#1\endcsname\relax
\typeout{** WARNING: IEEEtran.bst: No hyphenation pattern has been}%
\typeout{** loaded for the language `#1'. Using the pattern for}%
\typeout{** the default language instead.}%
\else
\language=\csname l@#1\endcsname
\fi
#2}}
\providecommand{\BIBdecl}{\relax}
\BIBdecl

\bibitem{Dimakis10}
A.~G. Dimakis, P.~B. Godfrey, Y.~Wu, M.~J. Wainwright, and K.~Ramchandran,
  ``Network coding for distributed storage systems,'' \emph{IEEE Trans. Inform.
  Theory}, vol.~56, no.~9, pp. 4539--4551, 2010.

\bibitem{Rashmi11}
K.~V. Rashmi, N.~B. Shah, and P.~V. Kumar, ``Optimal exact-regenerating codes
  for distributed storage at the {MSR} and {MBR} points via a product-matrix
  construction,'' \emph{IEEE Trans. Inform. Theory}, vol.~57, no.~8, pp.
  5227--5239, 2011.

\bibitem{Tamo13}
I.~Tamo, Z.~Wang, and J.~Bruck, ``Zigzag codes: {MDS} array codes with optimal
  rebuilding,'' \emph{IEEE Trans. Inform. Theory}, vol.~59, no.~3, pp.
  1597--1616, 2013.

\bibitem{Ye16}
M.~Ye and A.~Barg, ``Explicit constructions of high-rate {MDS} array codes with
  optimal repair bandwidth,'' \emph{IEEE Trans. Inform. Theory}, vol.~63,
  no.~4, pp. 2001--2014, 2017.

\bibitem{Sasid16}
B.~Sasidharan, M.~Vajha, and P.~V. Kumar, ``An explicit, coupled-layer
  construction of a high-rate {MSR} code with low sub-packetization level,
  small field size and all-node repair,'' 2016, arXiv:1607.07335.

\bibitem{Ye16a}
M.~Ye and A.~Barg, ``Explicit constructions of optimal-access {MDS} codes with
  nearly optimal sub-packetization,'' \emph{IEEE Trans. Inform. Theory},
  no.~10, pp. 6307--6317, 2017.

\bibitem{Tamo17RS}
I.~Tamo, M.~Ye, and A.~Barg, ``Optimal repair of {R}eed-{S}olomon codes:
  {A}chieving the cut-set bound,'' in \emph{Proc. 58th IEEE Sympos. on the
  Foundations of Computer Science (FOCS), October 15-17, 2017, Berkeley, CA},
  pp. 216--227.

\bibitem{Blaum98}
M.~Blaum, P.~G. Farell, and H.~van Tilborg, ``Array codes,'' in \emph{Handbook
  of Coding Theory}, V.~Pless and W.~C. Huffman, Eds.\hskip 1em plus 0.5em
  minus 0.4em\relax Elsevier Science, 1998, vol.~II, ch.~22, pp. 1855--1909.

\bibitem{Cadambe13}
V.~R. Cadambe, S.~A. Jafar, H.~Maleki, K.~Ramchandran, and C.~Suh, ``Asymptotic
  interference alignment for optimal repair of {MDS} codes in distributed
  storage,'' \emph{IEEE Trans. Inform. Theory}, vol.~59, no.~5, pp. 2974--2987,
  2013.

\bibitem{Rawat16a}
A.~S. Rawat, O.~O. Koyluoglu, and S.~Vishwanath, ``Centralized repair of
  multiple node failures with applications to communication efficient secret
  sharing,'' 2016, arXiv:1603.04822.

\bibitem{Wang17}
Z.~Wang, I.~Tamo, and J.~Bruck, ``Optimal rebuilding of multiple erasures in
  {MDS} codes,'' \emph{IEEE Transactions on Information Theory}, vol.~63,
  no.~2, pp. 1084--1101, 2017.

\bibitem{Zorgui17}
M.~Zorgui and Z.~Wang, ``Centralized multi-node repair for minimum storage
  regenerating codes,'' in \emph{2017 IEEE International Symposium on
  Information Theory (ISIT)}.\hskip 1em plus 0.5em minus 0.4em\relax IEEE,
  2017, pp. 2213--2217.

\bibitem{Ye17}
M.~Ye and A.~Barg, ``Repairing {R}eed-{S}olomon codes: {U}niversally achieving
  the cut-set bound for any number of erasures,'' 2017, arXiv:1710.07216.

\bibitem{Zorgui18}
M.~Zorgui and Z.~Wang, ``On the achievability region of regenerating codes for
  multiple erasures,'' 2018, arXiv:1802.00104.

\bibitem{Kermarrec11}
A.~M. Kermarrec, N.~Le~Scouarnec, and G.~Straub, ``Repairing multiple failures
  with coordinated and adaptive regenerating codes,'' in \emph{Network Coding
  (NetCod), 2011 International Symposium on}.\hskip 1em plus 0.5em minus
  0.4em\relax IEEE, 2011, pp. 1--6.

\bibitem{Shum13}
K.~W. Shum and Y.~Hu, ``Cooperative regenerating codes,'' \emph{IEEE
  Transactions on Information Theory}, vol.~59, no.~11, pp. 7229--7258, 2013.

\bibitem{Li14}
J.~Li and B.~Li, ``Cooperative repair with minimum-storage regenerating codes
  for distributed storage,'' in \emph{2014 Proceedings IEEE INFOCOM}.\hskip 1em
  plus 0.5em minus 0.4em\relax IEEE, 2014, pp. 316--324.

\bibitem{Shum16}
K.~W. Shum and J.~Chen, ``Cooperative repair of multiple node failures in
  distributed storage systems,'' \emph{International Journal of Information and
  Coding Theory}, vol.~3, no.~4, pp. 299--323, 2016.

\bibitem{Goparaju17}
S.~Goparaju, A.~Fazeli, and A.~Vardy, ``Minimum storage regenerating codes for
  all parameters,'' \emph{IEEE Trans. Inform. Theory}, vol.~63, no.~10, pp.
  6318--6328, 2017.

\bibitem{Elyasi16}
M.~Elyasi and S.~Mohajer, ``Determinant coding: {A} novel framework for
  exact-repair regenerating codes,'' \emph{IEEE Transactions on Information
  Theory}, vol.~62, no.~12, pp. 6683--6697, 2016.

\bibitem{Guruswami16}
V.~Guruswami and M.~Wootters, ``Repairing {R}eed-{S}olomon codes,'' \emph{IEEE
  Trans. Inform. Theory}, vol.~63, no.~9, pp. 5684--5698, 2017.

\bibitem{Dau17}
H.~Dau and O.~Milenkovic, ``Optimal repair schemes for some families of
  full-length {R}eed-{S}olomon codes,'' in \emph{Proc. 2017 IEEE International
  Symposium on Information Theory (ISIT)}.\hskip 1em plus 0.5em minus
  0.4em\relax IEEE, 2017, pp. 346--350.

\bibitem{Dau16}
H.~Dau, I.~Duursma, H.~M. Kiah, and O.~Milenkovic, ``Repairing {R}eed-{S}olomon
  codes with multiple erasures,'' 2016, arXiv:1612.01361.

\bibitem{Chowdhury17}
A.~Chowdhury and A.~Vardy, ``Improved schemes for asymptotically optimal repair
  of {MDS} codes,'' arXiv:1710.01867.

\bibitem{Bartan17}
B.~Bartan and M.~Wootters, ``Repairing multiple failures for scalar {MDS}
  codes,'' 2017, arXiv:1707.02241.

\bibitem{Ye16b}
M.~Ye and A.~Barg, ``Explicit constructions of {MDS} array codes and {RS} codes
  with optimal repair bandwidth,'' in \emph{Proc. 2016 IEEE International
  Symposium on Information Theory (ISIT)}.\hskip 1em plus 0.5em minus
  0.4em\relax IEEE, 2016, pp. 1202--1206.

\bibitem{Tamo18}
I.~Tamo, M.~Ye, and A.~Barg, ``The repair problem for {R}eed-{S}olomon codes:
  {O}ptimal repair of single and multiple erasures, asymptotically optimal node
  size,'' 2018, arXiv:1805.01883.

\bibitem{Cadambe11}
V.~R. Cadambe, C.~Huang, and J.~Li, ``Permutation code: Optimal exact-repair of
  a single failed node in {MDS} code based distributed storage systems,'' in
  \emph{Proc. 2011 IEEE Int. Sympos. Inform. Theory}, 2011, pp. 1225--1229.

\bibitem{BDMP98}
S.~Benedetto, D.~Divsalar, G.~Montorsi, and F.~Pollara, ``Serial concatenation
  of interleaved codes: performance analysis, design, and iterative decoding,''
  \emph{IEEE Transactions on Information Theory}, vol.~44, no.~3, pp. 909--926,
  1998.

\end{thebibliography}

\end{document}